\documentclass[reprint, notitlepage, amsmath, amssymb, onecolumn, pra]{revtex4-1} 
\usepackage[dvipsnames]{xcolor}
\usepackage{hyperref}
\hypersetup{
    colorlinks=true,
    linkcolor=black,
    citecolor=CadetBlue,
    filecolor=CadetBlue,      
    urlcolor=CadetBlue,
}

\pdfoutput=1 
\usepackage{graphicx}  
\usepackage{dcolumn}   
\usepackage{bm}        
\usepackage{amssymb}   
\usepackage{comment}
\usepackage{amsmath,amssymb,amsthm,dsfont,amsfonts,xfrac,braket,mathtools,bm,bbm,mathrsfs}

\newcommand{\pb}[2]{P_B^{#1}(\sigma_{#2})}
\newcommand{\q}[2]{Q(\sigma_{#2})^{#1}}

\newcommand{\graph}[1]{\mathcal{G}_{#1}}
\newcommand{\nbp}[3]{\mathcal{N}_{{#1}}(#2,#3)}

\usepackage{xcolor}

\usepackage{comment}

\usepackage[english]{babel}
\usepackage{amsmath,amsthm,amssymb}
\usepackage{amsfonts}
\usepackage{amsopn}
\usepackage{float}

\usepackage{url}

\usepackage{graphicx}
\usepackage{dcolumn}
\usepackage{bm}
\usepackage{color}

\usepackage{float}

\begin{document}
\author{Maria Chiara Angelini $^{1,2}$, Saverio Palazzi $^{1}$, Giorgio Parisi $^{1,2,3}$, Tommaso Rizzo $^{1,4}$}

\affiliation{$^1$Dipartimento di Fisica, Sapienza Universit\`a di Roma, Piazzale A. Moro 2, I-00185, Rome, Italy}
\affiliation{$^2$ INFN-Sezione di Roma 1, Piazzale A. Moro 2, 00185, Rome, Italy}
\affiliation{$^3$ Nanotec-CNR, UOS Rome, Sapienza Universit\`a di Roma, Piazzale A. Moro 2, I-00185, Rome, Italy}
\affiliation{$^4$ ISC-CNR, UOS Rome, Sapienza Universit\`a di Roma, Piazzale A. Moro 2, I-00185, Rome, Italy}

\title{Bethe \textit{M}-layer construction on the Ising model}

\begin{abstract}
In statistical physics, one of the standard methods to study second order phase transitions is the renormalization group that usually leads to an expansion around the corresponding fully connected solution. Unfortunately, often in disordered models, some important finite dimensional second-order phase transitions are qualitatively different or absent in the
corresponding fully connected model: in such cases the standard expansion fails. Recently, a new method, the $M$-layer one, has been introduced that performs an expansion around a different soluble mean field model: the Bethe lattice one. This new method has been already used to compute the upper critical dimension $D_U$ of different disordered systems such as the Random Field Ising model or the Spin glass model with field. If then one wants to go beyond and construct an expansion around $D_U$ to understand how critical quantities get renormalized, the actual computation of all the numerical factors is needed. This next step has still not been performed, being technically more involved. In this paper we perform this computation for the
ferromagnetic Ising model without quenched disorder, in finite dimensions: we show that, at one-loop order inside the $M$-layer approach, we recover the continuum quartic field theory and we are able to identify the coupling constant $g$ and the other parameters of the theory, as a function of macroscopic and microscopic details of the model such as the lattice spacing, the physical lattice dimension and the temperature. This is a fundamental step that will help in applying in the future the same techniques to more complicated systems, for which the standard field theoretical approach is impracticable. 

\end{abstract}


\maketitle

\tableofcontents

\hspace{3cm}

\section{Introduction}

Usually, the behavior of models in finite dimensions near a second-order phase transition can be deduced using the powerful method of Renormalization Group (RG) \cite{Parisi1988, amit2005field, Zinn-Justin_2002}. In field theory the basic mean-field (MF) approximation corresponds to the assumption that there are no fluctuations in the order parameter: this is equivalent to starting from the same model but defined on a fully connected topology, for which the global MF approximation is exact. Only when going beyond the leading order, one starts to add corrections computing the fluctuations of the order parameter in a $D$-dimensional underlying space. In the context of disordered systems, the quenched randomness of the interactions and/or fields causes physical observables to strongly depend on the local environment. This is one of the features that makes standard field theoretical predictions on these systems non-trivial \cite{FarBeyond_ch4_2023}. On the other hand, in the last decades, much effort has been devoted to the study of disordered models on finite-connectivity topologies. Among them, the locally tree-like graphs, also known as Bethe lattice (BL), play a crucial role. In fact, in a BL the marginal distribution for a given microscopic degree of freedom is independent of the probability of the nearest neighbors if the direct edge between them is cut: for this reason, a model on a BL is essentially mean-field in nature, and often solvable with the use of the so-called cavity method \cite{Mezard1987}, which coincides, at the Replica Symmetric (RS) level, to the ``Bethe-Peierls approximation'' \cite{bethe1935statistical}. Despite the MF nature, local observables on a BL do not coincide with global ones: local fluctuations and heterogeneities are possible because of the finite connectivity.
The difficulties in the analysis of finite-connectivity systems arise when one tries to include the contributions coming from the presence of topological loops of finite length. 

The $M$-layer construction is a recent attempt to systematically build a topological-loop expansion around the ``Bethe-Peierls approximation'' \cite{Altieri_2017}. In this framework it is possible to compute generic observables, starting from the value they take on a BL, and then perturbatively adding the contributions of topological loops.
To this aim, $M$ copies (called ``layers'') of the original lattice in $D$ dimensions are created. In this way, for each edge on the original lattice, there are $M$ couples of vertices. The second step of the construction consists of permuting uniformly at random the set of vertices, thereby generating inter-layer connections. Then it is possible to compute observables as if the initial model is defined on the topology resulting from one instance of the random rewiring of all the edges of the original lattice. It can be shown that the probability that loops are present in the resulting graph decreases as $M$ increases: in the $M\rightarrow\infty$ limit the resulting graph is locally tree-like. Assuming universality, for large but finite $M$, it is possible to use this construction to study the critical behavior of the original model (at $M=1$) without resorting to the standard field theory tool as a starting point: the result is a perturbative expansion for each chosen observable using $1/M$ as a small parameter where the leading term is obtained by computing the value that the chosen observable takes on a Bethe lattice. 

The $M$-layer construction has already been applied to many problems of interest, such as the random field Ising model \cite{Angelini_2019}, the spin glass in a field both at finite temperature in the limit of large connectivity \cite{angelini2018one} and at zero temperature and finite-connectivity \cite{Angelini_2021}, the Anderson localization \cite{baroni_2023}, the $k$-core percolation \cite{Rizzo_2019} and super-cooled liquids \cite{Rizzo_2020}. The majority of these problems could not be treated easily with standard field theory, either because they do not have a defined Hamiltonian or because, like in the case of Anderson localization, a transition is not possible in a fully connected topology, where there is not the possibility to define a distance between sites. In most of these non-trivial applications, the $M$-layer construction has been applied to understand which is the upper critical dimension, $D_U$, below which the MF description should be corrected. For determining $D_U$, it is sufficient to look at the divergences of the one-loop corrections to the leading behavior. 
If then one wants to go beyond and construct a true $\epsilon$-expansion around $D_U$ to understand how critical quantities get renormalized below $D_U$, the actual computation of all the numerical factors is needed. This next step has still not been applied, being technically more involved than the determination of $D_U$.
We believe that it could be really useful to perform this computation ``down to the metal'' for a simple model in which familiar results from the standard field-theoretical expansion should be recovered: the ferromagnetic Ising model without quenched disorder, in finite dimensions. This is the aim of this paper in which we will show that, at one-loop order, it is possible to recover the continuum $g\phi^4$ field theory from this new perspective, identifying the coupling constant $g$ and the other parameters of the theory, as a function of macroscopic quantities and microscopic details of the model: the lattice spacing, the physical lattice dimension, the temperature... This is a fundamental step that will help in applying the same techniques to more complicated systems in the future.
For the case of the ferromagnetic Ising model, we expect that an expansion around the Bethe lattice model, or around a fully connected model will give the same results because the ferromagnetic transitions have the same features on the two MF models. We in fact prove this, obtaining inside the $M$-layer construction the same perturbative corrections obtained inside the standard $g \phi^4$ field theory for the 2 and 4-point correlation functions. However, for more complex systems, such as the spin-glass with an external field, we do not expect the two expansions to give the same results, thus it is really important to validate the method on a test model such as the ferromagnetic Ising model, before proceeding towards more complicated, unpredictable models. 

The paper is organized as follows: In Sec. \ref{sec:modelmainresults} we introduce the model, we set the notation and we present the main results of the application of the $M$-layer construction to the Ising model. 
In Sec. \ref{sec:mlayer} we give a methodological overview of the $M$-layer construction. Furthermore, we provide all the building blocks to evaluate the 2 and 4-point connected correlation functions. As a working example, we compute some of the contributions to the 2-point connected correlation function. All the details of the computations for the 4-point observable can be found in Appendix \ref{app:mlayerconstruction}. Finally, we resume all the results in Sec. \ref{sec:conclusion}. Together with some comments, in Appendix \ref{app:caveats} we generalize the results of the main text, focused on Ising variables, to generic soft spin variables. 

\section{Model and main results}\label{sec:modelmainresults}
In this section, we present the results of the application of the $M$-layer construction to the Ising model in $D$ dimensions. Starting from the definition of the model, we describe in broad terms the steps to get the final result, which will be finally reported.

\subsection{The model}

The model is defined on the $D$-dimensional cubic lattice, with $D>1$, by the following Hamiltonian:
\begin{equation}\label{eq:ising_hamiltonian}
    H = - \sum_{i < j}^{1,\dots,N} J_{ij} \, \sigma_i \, \sigma_j\,,
\end{equation}
where $J_{ij}=1$ if $\sigma_i$ and $\sigma_j$ are nearest neighbours, $J_{ij}=0$ otherwise. Here we consider $N$ spin variables, $\sigma_i=\pm1$ $i=1,\dots,N$. The sites are identified by the positions $x_i\in a\mathbb{Z}^D$, where $a\mathbb{Z}^D$ denotes the $D$-dimensional cubic lattice with \emph{lattice spacing} ``\textit{a}'', which is the physical (dimensional) distance between nearest neighbors. We will apply the $M$-layer expansion to the Ising model and we will see that the results will be completely equivalent to a standard continuum $g\phi^4$ field theory. In this case, the field will correspond to the spin values in the lattice $\sigma:a\mathbb{Z}^D\mapsto\mathbb{R}$, in such a way that $\sigma(x_i)$ stands for $\sigma_i$.

\subsection{Results}

Here we report the results of the application of the $M$-layer machinery to the Ising model. We are interested in computing the critical behavior of the 2-point and 4-point connected correlation functions. They are defined as $\overline{\langle\sigma_i\sigma_j\rangle_c}=\overline{\langle\sigma_i\sigma_j\rangle}-\overline{\langle\sigma_i\rangle\langle\sigma_j\rangle}$ and $\overline{\langle\sigma_i\sigma_j\sigma_k\sigma_l\rangle_c}=\overline{\langle\sigma_i\sigma_j\sigma_k\sigma_l\rangle}-\overline{\langle\sigma_i\sigma_j\rangle\langle\sigma_k\sigma_l\rangle}-\overline{\langle\sigma_i\sigma_k\rangle\langle\sigma_j\sigma_l\rangle}-\overline{\langle\sigma_i\sigma_l\rangle\langle\sigma_k\sigma_j\rangle}$ respectively, where we denote with $\langle \cdot \rangle$ the thermal average and with $\overline{\, \cdot \,}$ the average over the disorder induced by the $M$-layer construction. The derivation of the 2-point function is provided in Sec. \ref{sec:2pointconnectedcorrelation}, while the derivation of the 4-point function in Appendix \ref{app:4pointconnectedcorrelation}.
 
Our main result is that the 2-point function, in Fourier space and in the large-length (small momenta) limit reads:

\begin{equation}\label{eq:2point}
    \overline{\langle\sigma(p)\sigma(q)\rangle_c} = (2\pi)^D\delta^D(p+q)  \,\frac{1}{\rho p^2+\tau} \Bigg(1  -\frac{1}{2} g \frac{1}{\rho q^2+\tau}\int_{[-\frac{\pi}{a},\frac{\pi}{a}]^D} \frac{d^Dk}{(2\pi)^D}\frac{1}{\rho k^2+\tau} \Bigg) +\mathcal{O}\left(\frac{1}{M^3}\right) \, ,
\end{equation}

where
\begin{align}
    & \rho = M \frac{a^{2-D}(1+\lambda)}{(2D)^4\lambda(1-\lambda)}\,  \label{eq:DefRho} \\[5pt]
    & \tau= M \frac{1}{(2D)^3\,a^D\,\lambda} \,  \, \Big( 1-\frac{\lambda}{\lambda_c} \Big)\label{eq:DefTau}\\[5pt]
    & g = 2 M\frac{(2D)!}{(2D-4)!}\,\frac{1}{(2D)^4\,a^D}\,  \label{eq:Defg}\\[5pt]
    & \lambda_c \equiv \frac{1}{2D-1} \label{eq:DefLambdac}\,.
\end{align}

The same can be done for the 4-point function: 

\begin{align}\label{eq:4point}
    &\overline{\langle\sigma(k_1)\sigma(k_2)\sigma(k_3)\sigma(k_4)\rangle_c}= \nonumber \\
    &\hspace{1.5cm}=\,(2\pi)^D\,\delta^D\left(\sum_{i=1}^4 k_i\right)\,\prod_{i=1}^4 \,G(k_i)\left[ -g +\frac{1}{2}g^2 \Big(I(k_1+k_2)+ I(k_1+k_3)+ I(k_1+k_4)\Big) \right]  +\mathcal{O}\left(\frac{1}{M^5}\right)\,,
\end{align}
where
\begin{align}
    I(q)&\equiv \int_{[-\frac{\pi}{a},\frac{\pi}{a}]^D} \frac{d^Dp}{(2\pi)^D}\frac{1}{\rho p^2+\tau}\cdot\frac{1}{\rho(p+q)^2+\tau}\,,\\
    G(q)&\equiv \frac{1}{\rho q^2+\tau} \Bigg(1  -\frac{1}{2} g \frac{1}{\rho q^2+\tau}\int_{[-\frac{\pi}{a},\frac{\pi}{a}]^D} \frac{d^Dk}{(2\pi)^D}\frac{1}{\rho k^2+\tau} \Bigg)\,.
\end{align}

In standard field theory, the Ising model is described by a Lagrangian for a field $\sigma$ defined on a continuum space with a quartic coupling interaction of the type:
\begin{equation}
\mathcal{L}=\frac{1}{2}\left[\rho(\nabla \sigma)^2+\tau\sigma^2\right]+\frac{1}{4!}g \sigma^4.
\label{eq:Lagrangian}
\end{equation}
If one computes the 2 and 4-point connected correlation functions associated with this Lagrangian expanding up to the first order in the coupling constant $g$, one obtains exactly Eqs. \eqref{eq:2point} and \eqref{eq:4point}. However, let us underline some conceptual differences between the standard field theory and the $M$-layer approach:
\begin{itemize}
    \item while in the standard field theory the parameters $\rho, \tau, g$ are usually introduced for symmetry reasons in the Lagrangian Eq. (\ref{eq:Lagrangian}), in the $M$-layer approach they are related, through Eqs. \eqref{eq:DefRho}, \eqref{eq:DefTau} and \eqref{eq:Defg}, to the physical quantities of the microscopic model: the dimension $D$, the lattice spacing $a$, and the leading eigenvalue $\lambda$ of the transfer-matrix that implicitly depends on the temperature $T$ (see Appendix \ref{app:transfermatrix} for more details). This connection between macroscopic and microscopic parameters is a simple exercise for the Ising model \cite{amit2005field}, but the $M$-layer construction allows it for a generic model, even if it is defined without the Hamiltonian, as in the case of the percolation problem; 
    \item while in the standard field theory the cutoff on the momentum integration is inserted manually, in the $M$-layer approach it naturally arises, being a function of the lattice spacing $a$, as can be seen from the computations, specifically from the theory on Non-Backtracking paths, see App. \ref{app:dimensionalanalysis}; 
    \item in standard field theory, the expansions in Eqs. \eqref{eq:2point} and \eqref{eq:4point} are perturbative in the dimensionless parameter associated with $g$, which is supposed to be small. In the $M$-layer approach $g, \rho, \tau$ are not small, being $\mathcal{O}(M)$, but using Eqs. \eqref{eq:DefRho}, \eqref{eq:DefTau} and \eqref{eq:Defg} it is easy to verify that the expansions for the correlation functions are actually perturbative in the parameter $1/M$ for $M\rightarrow\infty$, see Eq. \eqref{eq:correlation_shift} below. In particular, for the 2-point function the leading term is $\mathcal{O}\left(1/M\right)$ and the one-loop correction is $\mathcal{O}\left(1/M^2\right)$, while for the 4-point function the leading term is $\mathcal{O}\left(1/M^3\right)$ and the one-loop correction is $\mathcal{O}\left(1/M^4\right)$.
\end{itemize}

To make the perturbative nature of the $M$-layer expansion explicit, one should build a dimensionless parameter from the three dimensional parameters $g, \rho, \tau$. The only two dimensional initial quantities are the lattice spacing $a$, which has the dimension of a length, $[\,L\,]$, and the spins that have dimension of the field, $[\,\sigma\,]$. In appendix \ref{app:dimensionalanalysis} we show how to extract the dimensions of $g, \rho, \tau$, that turns out to be:
\begin{align}
    & [\,\tau\,]=[\,L\,]^{-D}[\,\sigma\,]^{-2}\\[5pt]
    & [\,\rho\,]=[\,L\,]^{2-D}[\,\sigma\,]^{-2}\\[5pt]
    & [\,g\,]=[\,L\,]^{-D}[\,\sigma\,]^{-4}\,,
\end{align}
Starting from $g, \rho, \tau$, the only dimensionless combination of them (with one power of $g$) is: $g\,\rho^{-\frac{D}{2}}\,\tau^{\frac{D-4}{2}}$. Using Eqs. \eqref{eq:DefRho}, \eqref{eq:DefTau} and \eqref{eq:Defg} is simple to verify that
\begin{equation}
    g\,\rho^{-\frac{D}{2}}\,\tau^{\frac{D-4}{2}} \propto \frac{1}{M}\,,
\end{equation}
which turns out to be proportional to $1/M$: we have proved that the only existing dimensionless combination of parameters is indeed small in the limit $M\to\infty$, verifying the perturbative nature of the previous expansion.
In this perspective, one can compute the shift of the critical temperature at $\mathcal{O}(1/M)$, redefining the constants $\tau\equiv M \tau'$, $\rho\equiv M \rho'$ and $g\equiv M g'$ and the 2-point correlation function:
\begin{equation}\label{eq:correlation_shift}
    M\,C(p)\equiv  \,\frac{1}{\rho' p^2+\tau'} \Bigg(1  -\frac{1}{2M} g' \frac{1}{\rho' p^2+\tau'}\int_{[-\frac{\pi}{a},\frac{\pi}{a}]^D} \frac{d^Dk}{(2\pi)^D}\frac{1}{\rho' k^2+\tau'} \Bigg)\,.
\end{equation}
If we consider $ M\,C(p=0)$ we see that the $1/M$ correction in the RHS is of order $1/\tau'$ and therefore it diverges at the critical temperature of the Bethe lattice, i.e. when $\tau'\to0$. To overcome this problem we follow the standard procedure of considering the inverse of $M C(p)$:
\begin{equation}
    (M\,C(p))^{-1}\equiv  \rho' p^2+\tau'    +\frac{1}{2M} g' \int_{[-\frac{\pi}{a},\frac{\pi}{a}]^D} \frac{d^Dk}{(2\pi)^D}\frac{1}{\rho' k^2+\tau'} \,.
\label{eq:MCm1}
\end{equation}
The condition $ (M\,C(0))^{-1}=0$ determines the critical value $\tau_c'$ that is zero at leading order in $1/M$:
\begin{align}
    &\tau_c' \equiv -\frac{1}{2M}g'\int_{[-\frac{\pi}{a},\frac{\pi}{a}]^D} \frac{d^Dk}{(2\pi)^D}\frac{1}{\rho'k^2}\label{eq:TempShift}\,.
\end{align}
Then we can consider the following expansion of $(M C(p))^{-1}$ for $p$ and $\tau'-\tau_c'$ both small:
\begin{equation}
    (M C(p))^{-1} = A(\tau'-\tau_c')+B p^2 +\mathcal{O}(p^4) +\mathcal{O}((\tau'-\tau_c')^2)\, .
\label{eq:MFexp}
\end{equation}
Comparing with Eq. \eqref{eq:MCm1} we obtain:
\begin{align}
    &A=1-\frac{1}{2M}g'\int_{[-\frac{\pi}{a},\frac{\pi}{a}]^D} \frac{d^Dk}{(2\pi)^D}\frac{1}{(\rho'k^2)^2}\\
    & B=\rho'\,.
\end{align}
The expression for $\tau_c'$ is divergent for $D \leq 2$ and the expression for $A$ is divergent for $D \leq 4$, signaling that the MF expansion on Eq. (\ref{eq:MFexp}) is only valid above $D=4$ that is indeed the upper critical dimension of the problem.




\section{The $M$-layer construction}\label{sec:mlayer}

In this section, we set the stage to describe, step by step, the $M$-layer construction. We will start with some general features and then we will analyze the computation of some contributions to the 2-point connected correlation function for the Ising model.

The $M$-layer procedure consists in creating $M$ copies (layers) of the original lattice, on which the system is defined. For each edge, connecting $i$ and $j$ in the original lattice, there are $M$ copies $(i_1,j_1)$, $(i_2,j_2)$, $\dots$, $(i_M,j_M)$. To create inter-layer connections, a permutation between the two sets of vertices, $(i_1,i_2,\dots,i_M)$ and $(j_1,j_2,\dots,j_M)$, is chosen uniformly at random between the $M!$ possible ones, rewiring the sites. Repeating this procedure for all the $M$ copies of all the edges belonging to the original lattice, one can show that in the limit $M\rightarrow\infty$ the resulting graph is locally tree-like. The details of this statement can be found in \cite{Altieri_2017}. Intuitively, the probability to close a short loop is proportional to $1/M$, thus a tree-like graph is realized for $M\to\infty$. The average over all the possible rewirings has to be taken at the end of the computation. With this construction and considering the large-$M$ limit, it is possible to compute a generic observable on the layered lattice in a perturbative fashion, where the infinitesimal parameter is $1/M$. In the leading order the observable as computed on a tree graph, corresponding to the Bethe-Peierls approximation. The next order corrections correspond to adding loops to the system. 

\subsection{General features of the construction: the example of the Ising model}
\label{sec:rules}
Let us consider the Ising model on the $a\mathbb{Z}^D$ lattice and apply the $M$-layer construction. We want to compute a specific observable: the 2-point connected correlation function between two spins averaged over the rewirings. To do so, a way to organize the sum of the different contributions is needed. Then, considering a specific instance of the construction, we can use the cavity method to compute the chosen observable. In order to do so we will use the definition of ``line-connected observable'' following the guidelines of the $M$-layer construction \cite{Altieri_2017}.

 Since we are considering the connected correlation, we expect it to be zero if the graph resulting from the rewiring is such that no path connects the two spins. On the other hand, the correlation would be non-null if the chosen rewiring connects the two spins with a single loop-less path, a line, or with paths that include loops. 

We notice that, given the path type, the observable depends only on the lengths of the lines, therefore we can simply consider the projection of the lines on the original lattice. With this remark, the whole sum can be drastically eased: we sum over the projected paths with the corresponding weight $W$, that is the number of the rewirings that share the same projection, divided by $M!^{|E|}$, where $|E|$ is the number of edges.  It is possible to classify the projections by their topologies. We distinguish different classes by the presence of loops and internal vertices of a specific degree, that is the number of lines entering the vertex. We will refer to these classes as “diagrams”, to make the connection with the Feynman diagrams of field theory, with the important difference that topological lines make the former, whereas the latter are a graphic way to compute integrals. In the $M$-layer construction, given a diagram $\mathcal{G}$ projected onto the original lattice, each line corresponds to a Non-Backtracking Path (NBP) between a starting and an ending point. The number of NBPs of a fixed length $L$, between two points, $x$ and $y$, in the cubic lattice has already been computed \cite{Fitzner_2013}, we will denote it with $\mathcal{N}_L(x,y)$. At this point we recap the general steps to compute a generic observable averaged over the rewirings, following ref. \cite{Altieri_2017}:

\begin{itemize}
    \item Identify the possible diagrams (classes of projections on the original lattice) that contribute to the observable;
    \item For each diagram $\mathcal{G}$, compute its contribution as follows: 
    \begin{itemize}
        \item Compute the corresponding weight $W(\mathcal{G})$;
        \item Multiply the precedent term by a factor $\mathcal{N}_{L}(x,y)$ for each line, with length $L$, composing the diagram, starting at $x$ and ending at $y$;
        \item Multiply the precedent term by the value of the line-connected observable on a Bethe lattice in which the identified diagram has been manually injected, computed using the RS cavity method;  
        \item Sum over the positions of the internal vertices and over the lengths of the lines;
    \end{itemize} 
    \item Sum the contributions coming from all diagrams;
    \end{itemize}

Once the diagrams are identified, it is clear that the three building blocks are the weight, the number of NBPs and the line-connected observable computed on the diagram. While the first two are model-independent and well-understood, the third is strongly dependent on the model, thus it must be computed case by case.

\subsection{An example: the 2-point connected correlation function}\label{sec:2pointconnectedcorrelation}
In this paragraph, we will show the computations for the 2-point function, made by the following steps.

$\bullet$ \textit{Identify the possible diagrams:} 

\begin{figure}[H]
    \centering
    \includegraphics[scale=0.4]{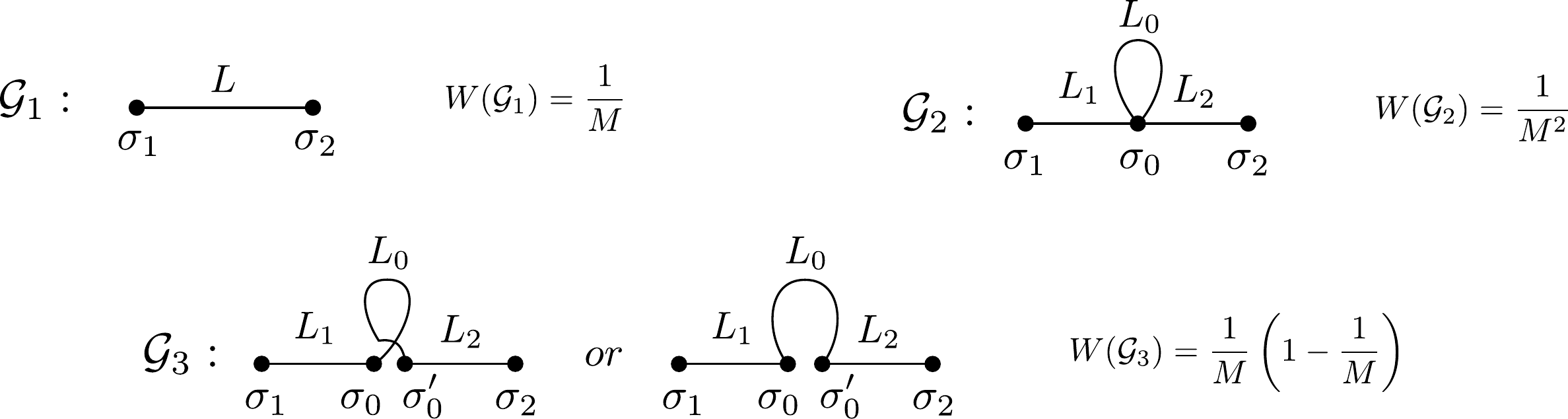}
    \caption{Topological diagram contributions to the 2-point connected correlation function. Top left: the order $\mathcal{O}(1/M)$, that is the simple line without loops in the layered lattice nor in the original one. Top right: the diagram with one loop is of the order $\mathcal{O}(1/M^2)$. Bottom: the so-called ``\textit{boucle}'' diagram, whose weight is $(1/M-1/M^2)$ where the line doesn't close in a loop on the layered lattice but realizes it in the projection.}
    \label{fig:2point}
\end{figure}

Following the general rules, the starting point is to identify the possible diagrams that contribute to the chosen observable, that is the 2-point connected correlation function. The simplest diagrams are obtained when we connect two points with a single loopless line, of length $L$, in the layered lattice. Given that the $1/M$ expansion turns out to be an expansion in the number of topological loops, the contribution of this type of diagrams will give the leading order in the expansion \footnote{In general, for 2-point observables that are not connected, such for example the total correlation function, additional diagrams in which the 2-points are disconnected should be considered, too. However, for connected observables such as the correlation functions the contribution of these diagrams is null.}.
Now we should understand which type of projection on the original lattice should be generated by a simple line in the layered lattice. The simplest projection is a simple loopless line also on the original lattice: we will call this diagram $\mathcal{G}_1$, and it is reported in Fig. \ref{fig:2point}.
However it is possible that a single line on the layered lattice will turn to be a line with loops on the original lattice: we will call the diagram with one loop on the original lattice as $\mathcal{G}_3$, it is reported in Fig. \ref{fig:2point} in its two versions and we will refer to them as ``\textit{boucle}'' diagrams.
 Another possibility is to connect the two spins with a path with one loop already on the layered lattice, which thus will have (at least) one loop also on its projection on the original lattice: this is the case of $\mathcal{G}_2$ in Fig. \ref{fig:2point}. $\mathcal{G}_2$ has a 4-degree topological vertex; in principle, we should consider also one-loop diagrams with 3-degree vertices. However, by introducing the RS cavity method to compute the line-connected observable on a given diagram, we will understand that they would give zero contribution for symmetry reasons in the case of the Ising model without an external field.

$\bullet$ \textit{Compute the contribution for diagrams $\mathcal{G}_1$ and $\mathcal{G}_3$:}

We start by computing the corresponding weight $W(\mathcal{G}_1)$, that is the probability of generating a 
rewiring that connects the two spins $\sigma_1$ and $\sigma_2$ with a unique path, without making loops in the original (projected) lattice. 
Let us start from the position of the first spin $\sigma_1$. The first edge can be chosen to connect $\sigma_1$ with the following spin placed at any of the possible $M$ layers, and each layer is chosen with probability $1/M$, so that the total contribution of the first edge to $W(\mathcal{G}_1)$ is 1. The same is true for all the edges up to the last one, which connects to the ending spin $\sigma_2$. Since the position of $\sigma_2$ is also fixed we have only 1 possible edge to choose, while the probability of choosing that specific edge is still $1/M$, thus $W(\mathcal{G}_1)=1/M$.

Let us pass now to the weight $W(\mathcal{G}_3)$, that is the probability of generating a 
rewiring that connects the two spins $\sigma_1$ and $\sigma_2$ with a unique linear path with no loops in the layered lattice, while the path projected on the original lattice has one loop when the path coils itself on a spin $\sigma_0$. As for $\mathcal{G}_1$, the first edges give a contribution 1, until one reaches the edge that arrives at $\sigma_0$ for the second time: to avoid creating a loop on the layered lattice, one should only pick the $\sigma_0$ on layers that were not previously visited. They will be $M-1$, and each of them is chosen with probability $1/M$, giving a factor $(1-1/M)$. As for $\mathcal{G}_1$, also the choice of the last edge ending in $\sigma_2$ gives an additional factor $1/M$: thus we find $W(\mathcal{G}_3)=1/M-1/M^2$. $W(\mathcal{G}_3)$, as computed here, is the weight of a linear path that produces only one loop on the original lattice. But now we have understood how to generalize the previous computation: for a linear path on the layered lattice, we should add a factor $(1-1/M)$ for each spin visited twice on the original lattice. For example, a path with no loops on the layered lattice and two loops on the projected original one will have a weight $1/M(1-1/M)^2$. 
Thus the leading order for the weight of all these diagrams will be $\mathcal{O}(1/M)$, and the presence of loops in the original lattice will only enter as a next-to-leading order correction for $W$.
Following the prescriptions of Sec. \ref{sec:rules}, the weight of each diagram $\mathcal{G}$ should then be multiplied by $\langle\sigma_1\sigma_2\rangle_c \Big|_{\mathcal{G},\text{lc}}$, that is the 2-point connected correlation function computed in a Bethe lattice where $\sigma_1$ and $\sigma_2$ are connected by a diagram whose projection is $\mathcal{G}$ (and the subscript ``lc'' stands for line-connected).
However, it is easy to understand that $\langle\sigma_1\sigma_2\rangle_c \Big|_{\mathcal{G},\text{lc}}$ just depends on the form of $\mathcal{G}$ on the layered lattice, and not on its projection on the original lattice. This observation implies that $\langle\sigma_1\sigma_2\rangle_c \Big|_{\mathcal{G}_1,\text{lc}}=\langle\sigma_1\sigma_2\rangle_c \Big|_{\mathcal{G}_3,\text{lc}}$ and in general it would be the same for any diagram, of the same length $L$, that has no loops on the layered graph, regardless of the number of loops on the projected lattice.

The last ingredient to take into account is the number of realizations of such diagrams on the original lattice. In principle it is possible to count the number of NBPs of length $L$ between two points, $x$ and $y$ in $a\mathbb{Z}^D$, specifying the starting and ending directions, that we call $\mu$ and $\nu$ respectively, denoted by $\mathcal{N}_L(x,y;\mu,\nu)$. This number will include the linear paths without loops, corresponding to diagrams $\mathcal{G}_1$ but also the paths with one loop, corresponding to diagrams $\mathcal{G}_3$ and in general diagrams with an arbitrary number of loops on the projected $D$-dimensional lattice. Remembering that the leading order for the weight of all these diagrams is $1/M$, and that the observable computed on a diagram $\mathcal{G}$, $\langle\sigma_1\sigma_2\rangle_c \Big|_{\mathcal{G},\text{lc}}$, is the same for all the diagrams obtained as a projection of a linear path on the layered lattice, we can write the first contribution to the 2-point connected correlation function:
\begin{equation}
    \overline{\langle\sigma(x_1)\sigma(x_2)\rangle_c}=\frac{1}{M} \sum_{\mu,\nu}\sum_{L=1}^{\infty} \mathcal{N}_L(x_1,x_2;\mu,\nu) \, \langle\sigma_1\sigma_2\rangle_c \Big|_{\mathcal{G}_1,\text{lc}} + \mathcal{O}(1/M^2)\simeq\frac{(2D)^2}{M} \sum_{L=1}^{\infty} \mathcal{N}_L(x_1,x_2) \, \langle\sigma_1\sigma_2\rangle_c \Big|_{\mathcal{G}_1,\text{lc}}+ \mathcal{O}(1/M^2)\,.
    \label{eq:2point_O1/M}
\end{equation}
In the last step, we assumed that the number of NBPs doesn't depend on the specific directions $\mu$ and $\nu$, thus we simply multiplied by the number of all the possible starting and ending directions, $(2D)^2$. This last assumption is justified if the lengths of the lines diverge, so it is justified in the critical region, where the most important contributions to the correlation functions come from the large-length lines. Summing over the directions, it is possible to over-count the possible paths; in that case, we must divide by a symmetry factor, associated with each diagram $\mathcal{G}$, which we call $S(\mathcal{G})$. In the case of $\mathcal{G}_1$, we are not over-counting the paths, so $S(\mathcal{G}_1)=1$.

We have said that diagrams $\mathcal{G}_3$ have a weight $1/M-1/M^2$ and in general the weights of all diagrams corresponding to linear paths with at least one loop in the projected graph have a $1/M^2$ correction. Thus, to compute the $\mathcal{O}(1/M^2)$ correction to Eq. \eqref{eq:2point_O1/M}, we should count all the possible linear paths that have a projection with at least one loop, the first example being $\mathcal{G}_3$. The $1/M^2$ correction due to linear-path diagrams is thus:

\begin{equation}
    -\frac{(2D)^2}{M^2} \frac{(2D)!}{(2D-4)!}\sum_{x_0\in a\mathbb{Z}^D}\sum_{L_0,L_1,L_2=1}^{\infty} \mathcal{N}_{L_1}(x_1,x_0)\mathcal{N}_{L_0}(x_0,x_0)\mathcal{N}_{L_2}(x_0,x_2) \, \langle\sigma_1\sigma_2\rangle_c \Big|_{\mathcal{G}_3,\text{lc}}\,,
    \label{eq:2point_O1/M2}
\end{equation}
where the $-$ sign comes from the sign of $1/M^2$ into $W$. We have multiplied the number of NBPs for each line and we have an additional factor due to the presence of (at least) one internal 4-degree vertex, where the sum over the four directions accounts for the number of ordered 4-uples out of the 2D possible directions: $\binom{2D}{4}4!$, exploiting the assumption of independence of $\mathcal{N}_L$ on the directions of the lines entering the vertices. In general, for a $k$-degree vertex, the factor will be $\binom{2D}{k}k!$. The symmetry factor is 1 also in this case, $S(\mathcal{G}_3)=1$. This can be understood since we are counting all the possible paths that coil to realize a projection with (at least) one loop, thus exchanging the two directions of the lines of the loop entering the internal vertex gives two different possibilities we should take into account, as shown in Fig. \ref{fig:2point}. As already mentioned, $\langle\sigma_1\sigma_2\rangle_c \Big|_{\mathcal{G}_3,\text{lc}}$ has to be equal to $\langle\sigma_1\sigma_2\rangle_c \Big|_{\mathcal{G}_1,\text{lc}}$ changing $L$ into $L_1+L_0+L_2$ .
We want to stress once more that Eq. \eqref{eq:2point_O1/M2} contains the contribution coming from all the linear paths resulting in a projection with at least one loop. The diagrams with at least 2 loops in the projection will then bring an additional $\mathcal{O}(1/M^3)$ correction, that we will not compute in this paper.

$\bullet$ \textit{Compute the contribution for diagram $\mathcal{G}_2$:}

At this point we should add the contributions of the one-loop diagram $\mathcal{G}_2$ in Fig. \ref{fig:2point}, where the observable is computed in a graph where $\sigma_1$ and $\sigma_2$ are connected by a line with such a loop on the layered lattice:
\begin{equation}
    W(\mathcal{G}_2) \frac{(2D)^2}{S(\mathcal{G}_2)} \frac{(2D)!}{(2D-4)!} \sum_{x_0\in a\mathbb{Z}^D}\sum_{L_0,L_1,L_2=1}^{\infty} \mathcal{N}_{L_1}(x_1,x_0)\mathcal{N}_{L_0}(x_0,x_0)\mathcal{N}_{L_2}(x_0,x_2) \, \langle\sigma_1\sigma_2\rangle_c \Big|_{\mathcal{G}_2,\text{lc}}\,,
    \label{eq:G2}
\end{equation}
where $W(\mathcal{G}_2)=1/M^2$. In fact, to create a loop on the layered lattice, the second edge that visits $\sigma_0$ should only pick the exact layer that was previously visited, which is chosen with probability $1/M$, giving another factor $1/M$ to the weight. As for $\mathcal{G}_1$, also the choice of the last edge ending in $\sigma_2$ gives an additional factor $1/M$: thus we find $W(\mathcal{G}_2)=1/M^2$.

In this case $S(\mathcal{G}_2)=2$, since, when summing over the directions of the internal vertices, all the terms where the two directions of the line of the loop are exchanged give the same diagram, so we should divide by a factor 2.
Exactly as for $\mathcal{G}_1$, Eq. \eqref{eq:G2} for the contribution of one-loop diagrams should be corrected at order $1/M^3$ for the presence of one-loop diagrams with lines that create (at least) one more loop on the projected lattice.

$\bullet$ \textit{Sum the contributions for the three analyzed diagrams:} 

Now we sum up the contributions, leaving as a last step the explicit computation of the line-connected observable using the RS cavity method.

Up to order $1/M^2$ these are all the possible contributions:

\begin{align}\label{eq:2pointAverage}
    \overline{\langle\sigma(x_1)\sigma(x_2)\rangle_c} = & \frac{(2D)^2}{S(\graph{1})M} \sum_{L=1}^{\infty} \mathcal{N}_L(x_1,x_2) \, \langle\sigma_1\sigma_2\rangle_c \Big|_{\mathcal{G}_1,\text{lc}} + \nonumber\\
    &+ \frac{(2D)^2}{S(\graph{2})M^2} \frac{(2D)!}{(2D-4)!}\sum_{x_0\in a\mathbb{Z}^D}\sum_{L_0,L_1,L_2=1}^{\infty} \mathcal{N}_{L_1}(x_1,x_0)\mathcal{N}_{L_0}(x_0,x_0)\mathcal{N}_{L_2}(x_0,x_2) \, \langle\sigma_1\sigma_2\rangle_c \Big|_{\mathcal{G}_2,\text{lc}} \nonumber\\
    &- \frac{(2D)^2}{S(\graph{3})M^2}  \frac{(2D)!}{(2D-4)!}\sum_{x_0\in a\mathbb{Z}^D}\sum_{L_0,L_1,L_2=1}^{\infty} \mathcal{N}_{L_1}(x_1,x_0)\mathcal{N}_{L_0}(x_0,x_0)\mathcal{N}_{L_2}(x_0,x_2) \, \langle\sigma_1\sigma_2\rangle_c \Big|_{\mathcal{G}_3,\text{lc}}+\mathcal{O}\left(\frac{1}{M^3}\right)\,.
\end{align}

 We have seen that the term of order $1/M$ includes diagrams with one loop, whose contribution we have corrected by computing the \textit{boucle} diagram that gives a contribution at order $1/M^2$; exactly in the same way, the term of order $1/M^2$ includes diagrams with two or more loops in the layered lattice, which should therefore be subtracted. However, these latter contributions will be of order $1/M^3$, so we can neglect them in the calculation of the 2-point connected correlation function at one loop level. Another important remark is that the \textit{boucle} diagrams are not present for vertices of degree three. This can be easily verified from a simple geometrical point of view: it is not possible to coil a line on a point to create a 3-degree vertex on the projection, without overlapping an entire internal line, hence obtaining a correction of larger $1/M$ order. The \textit{boucle} diagrams were already considered in \cite{Altieri_2017}, arguing that they are generally negligible. We show here that in the case of the Ising model they are not negligible. However, all the subsequent works that used the $M$-layer expansion introduced in ref. \cite{Altieri_2017}, such as refs. \cite{Angelini_2019,angelini2018one,Angelini_2021,baroni_2023,Rizzo_2019,Rizzo_2020}, looked to diagrams with cubic vertices for which \textit{boucle}-diagrams are indeed negligible.

$\bullet$ \textit{Compute the line-connected observables through the RS cavity method:} 

The next step is to evaluate the observable on a diagram, merged in an infinite tree. At the end of the $M$-layer construction the graph is, in the large $M$ limit, locally tree-like, thus we want to make use of the Replica Symmetric Cavity Method \cite{Mezard1987,Yedidia2003}. On a BL, two nearest neighboring spins are independent once the link that connects them is cut. For this reason, in the case of the Ising model on a random regular graph, one can implicitly define the ``cavity marginal distribution'', $Q(\sigma)$, as the solution of the following equation:
\begin{align}
    Q(\sigma)&=\frac{1}{\mathcal{Z}_Q}\sum_{\tau=\pm1}\,e^{\beta \sigma \tau }Q^{c-1}(\tau) \label{eq:Qmain}\\
    \mathcal{Z}_Q&=\sum_{\tau,\sigma=\pm1}\,e^{\beta \sigma \tau }Q^{c-1}(\tau)\,,
\end{align}
where $\beta$ is the inverse temperature, $c$ is the fixed connectivity, $Q(\sigma)$ is the cavity distribution on spin $\sigma$ where only the contribution coming from one interaction is considered  and $\mathcal{Z}_Q$ ensures the normalization of $Q$. For symmetry reasons, it is easy to understand that the solution to the previous implicit equation is $Q(\pm1)=\frac{1}{2}$.

Now we consider an open chain of length $L$, merged in a tree-like graph where the two variables at the boundaries, which we denote with $\sigma$ and $\tau$, have connectivity 1. As shown in Appendix \ref{app:mlayerconstruction}, the (unnormalized) joint probability distribution of $\sigma$ and $\tau$, which we define $Z_L(\sigma,\tau)$, can be written as
\begin{equation}\label{eq:jointprobdistribution}
    Z_L(\sigma,\tau)=P_B(\sigma)P_B(\tau)+\lambda^L\,{g_{\lambda}(\sigma)\,g_{\lambda}(\tau)}\,,
\end{equation}
where $P_B(\pm1)=2^{\frac{c-3}{2}}$ and $g_{\lambda}(\pm1)=\pm2^{\frac{c-3}{2}}$ are eigenvectors of a properly defined $2\times2$ transfer matrix (see App. \ref{app:transfermatrix}), with eigenvalues 1 and $\lambda$ respectively. $\lambda$ is temperature-dependent and assumes the critical value, $\lambda_c=1/(c-1)$, at the critical temperature $T_c$, causing the divergence of the ferromagnetic susceptibility. Notice that $P_B(\sigma)$ and $g_{\lambda}(\sigma)$ are even and odd functions of $\sigma=\pm1$, which is a consequence of the $\mathbb{Z}_2$ symmetry of the Ising model.

Once $Z_L(\sigma,\tau)$ is computed we can evaluate a generic observable on a generic diagram, taking also into account the presence of loops. Given a diagram, we identify a spin on each internal and external vertex and we multiply a factor $Z_L(\sigma,\tau)$ for each line of length $L$, starting at spin $\sigma$ and ending at spin $\tau$. Next, remembering that each spin has fixed connectivity $c$, we multiply a factor $Q(\sigma_I)^{c-r}$ for each internal vertex of degree $r$ on spin $\sigma_I$, and a factor $Q(\sigma_E)^{c-1}$ for each external vertex with spin $\sigma_E$. We recall that $Q(\sigma)$, defined self-consistently in Eq. (\ref{eq:Qmain}), is the cavity marginal where only the contribution coming from one interaction is considered. At this point, to compute the 2-point correlation function we multiply by the two external spin variables and we integrate over all the spins with the corresponding measure. We must also divide by the partition function and subtract all the factorized contributions to compute the connected correlation. In the end, we should compute the ``line-connected observable'', following the prescriptions of the $M$-layer construction \cite{Altieri_2017}.

\begin{figure}[H]
    \centering
    \includegraphics[scale=0.8]{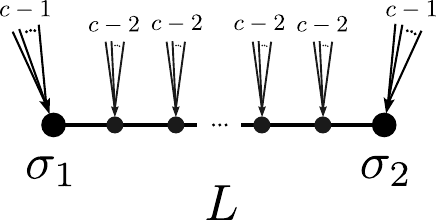}
    \caption{A diagram made of a simple line of length $L$, which we defined $\mathcal{G}_1$, embedded into a Bethe lattice is represented. On each of the $L-1$ internal spins, $c-2$ cavity fields are coming from an infinite tree graph, represented by the arrows, while the two spins at the boundaries are connected with $c-1$ of them. In this way, the diagram is merged into an infinite tree graph. More complicated diagrams are composed of such lines, but for simplicity we didn't include the arrows for Figs. \ref{fig:2point}, \ref{fig:4point} and \ref{fig:G5aG5b}. }
    \label{fig:G1}
\end{figure}

We start evaluating the 2-point connected correlation function for the system on a diagram $\mathcal{G}_1$ inserted in a Bethe lattice, as depicted in Fig. \ref{fig:G1}.

Remembering that $P_B(\sigma)$ and $Q(\sigma)$ are even functions while $g_{\lambda}(\sigma)$ is an odd one, we easily obtain:
\begin{equation}\label{eq:g1lc}
     \langle\sigma_1\sigma_2\rangle_c  \Big|_{\mathcal{G}_1,\text{lc}} = \frac{\sum\limits_{\sigma_1,\sigma_2=\pm1}  \,Q(\sigma_1)^{c-1}\,\sigma_1\,Z_L(\sigma_1,\sigma_2)\,\sigma_2\,Q(\sigma_2)^{c-1}}{\sum\limits_{\sigma_1,\sigma_2=\pm1}\,Q(\sigma_1)^{c-1}\,Z_L(\sigma_1,\sigma_2) \,Q(\sigma_2)^{c-1}} =\lambda^{L}\,\frac{\Big(\sum\limits_{\sigma=\pm1} \sigma\,g_{\lambda}(\sigma)\,Q(\sigma)^{c-1}\Big)^2}{\Big(\sum\limits_{\sigma=\pm1} P_B(\sigma)\,Q(\sigma)^{c-1}\Big)^2}=\lambda^{L}\,.
\end{equation}

Next, we move to the one-loop diagram $\mathcal{G}_2$.
In this case we have:
\begin{align}\label{eq:g2NONlc}
    \langle\sigma_1\sigma_2\rangle_c  \Big|_{\mathcal{G}_2} 
    & = \frac{\sum\limits_{\sigma_1,\sigma_0,\sigma_2=\pm1} \,Q(\sigma_1)^{c-1}\,\sigma_1\,Z_{L_1}(\sigma_1,\sigma_0)\,Q(\sigma_0)^{c-4}\,Z_{L_0}(\sigma_0,\sigma_0)\,Z_{L_2}(\sigma_0,\sigma_2)\,\sigma_2\,Q(\sigma_2)^{c-1}}{\sum\limits_{\sigma_1,\sigma_0,\sigma_2=\pm1}\,Q(\sigma_1)^{c-1}\,Z_{L_1}(\sigma_1,\sigma_0)\,Q(\sigma_0)^{c-4}\,Z_{L_0}(\sigma_0,\sigma_0)\,Z_{L_2}(\sigma_0,\sigma_2)\,Q(\sigma_2)^{c-1}} \nonumber\\
    & = \lambda^{L_1+L_2}\, \frac{\Big(\sum\limits_{\sigma=\pm1} \sigma\,g_{\lambda}(\sigma)\,Q(\sigma)^{c-1}\Big)^2}{\Big(\sum\limits_{\sigma=\pm1} P_B(\sigma)\,Q(\sigma)^{c-1}\Big)^2}\left( \frac{\sum\limits_{\sigma_0=\pm1} g_{\lambda}(\sigma_0)^2\,P_B(\sigma_0)^2\,Q(\sigma_0)^{c-4}+\lambda^{L_0}\sum\limits_{\sigma_0=\pm1} g_{\lambda}(\sigma_0)^4\,Q(\sigma_0)^{c-4}}{\sum\limits_{\sigma_0=\pm1} \,P_B(\sigma_0)^4\,Q(\sigma_0)^{c-4}+\lambda^{L_0}\sum\limits_{\sigma_0=\pm1} g_{\lambda}(\sigma_0)^2\,P_B(\sigma_0)^2\,Q(\sigma_0)^{c-4}}\right) \nonumber\\
    & =\lambda^{L_1+L_2}\,.
\end{align}

At this point, we can compute the ``line-connected'' 2-point connected correlation function. Let us try to make evident the need for such a ``connectification'' (the rigorous derivation can be found in \cite{Altieri_2017}). Take the diagram $\mathcal{G}_2$ as an example: In computing the 2-point correlation function, two terms will appear due to $Z_L(\sigma_0,\sigma_0)$, one with $P_B(\sigma_0)P_B(\sigma_0)$ and the second with $g_\lambda(\sigma_0)g_\lambda(\sigma_0)$. Since $P_B(\sigma)$ is proportional to $Q(\sigma)$, as explained in detail in Appendix \ref{app:transfermatrix}, the former is the same contribution as the one that is obtained from $\mathcal{G}_1$. Computing the line-connected observable will avoid this over-counting; the line-connected observable is thus defined as:
\begin{align}\label{eq:g2lc}
      \langle\sigma_1\sigma_2\rangle_c \Big|_{\mathcal{G}_2,\text{lc}} &=\langle\sigma_1\sigma_2\rangle_c \Big|_{\mathcal{G}_2} - \langle\sigma_1\sigma_2\rangle_c \Big|_{\mathcal{G}_1} = \nonumber\\
      &\lambda^{L_1+L_2}\,\frac{\Big(\sum\limits_{\sigma=\pm1} \sigma\,g_{\lambda}(\sigma)\,Q(\sigma)^{c-1}\Big)^2}{\Big(\sum\limits_{\sigma=\pm1} P_B(\sigma)\,Q(\sigma)^{c-1}\Big)^2} \left(\frac{\sum\limits_{\sigma=\pm1} g_{\lambda}(\sigma)^2\,P_B(\sigma)^2\,Q(\sigma)^{c-4}+\lambda^{L_0}\sum\limits_{\sigma=\pm1} g_{\lambda}(\sigma)^4\,Q(\sigma)^{c-4}}{\sum\limits_{\sigma=\pm1} \,P_B(\sigma)^4\,Q(\sigma)^{c-4}+\lambda^{L_0}\sum\limits_{\sigma=\pm1} g_{\lambda}(\sigma)^2\,P_B(\sigma)^2\,Q(\sigma)^{c-4}}-1\right)=0 \,,
\end{align}
where we used Eq. \eqref{eq:g1lc} with $L=L_1+L_2$.

The fact that $ \langle\sigma_1\sigma_2\rangle_c \Big|_{\mathcal{G}_2,\text{lc}}$ is trivially zero suggests that some more corrections must be added to the 2-point correlation function at $\mathcal{O}\left(1/M^2\right)$. As explained before, these corrections are due to the \textit{boucle} diagrams. $\langle\sigma_1\sigma_2\rangle_c \Big|_{\mathcal{G}_3,\text{lc}}$ is the same as the one computed on $\mathcal{G}_1$ but with the length of the path indicated as $L_1+L_0+L_2$:

\begin{equation}\label{eq:g3lc}
    \langle\sigma_1\sigma_2\rangle_c \Big|_{\mathcal{G}_3,\text{lc}} = \lambda^{L_1+L_0+L_2}\,. 
\end{equation}

At this point, we plug Eqs. \eqref{eq:g1lc}, \eqref{eq:g2lc} and \eqref{eq:g3lc} into Eq. \eqref{eq:2pointAverage}, and remembering that
\begin{align}
    &W(\graph{1})=\frac{1}{M}\,\qquad\qquad\qquad\quad S(\graph{1}) = 1 \qquad\qquad \langle\sigma_1\sigma_2\rangle_c\Big|_{\mathcal{G}_1,\,\text{lc}}= \lambda^{L}; \nonumber\\
    &W(\graph{2})=\frac{1}{M^2}\qquad\qquad\qquad\,\,\,\, S(\graph{2}) = 2 \qquad\qquad \langle\sigma_1\sigma_2\rangle_c\Big|_{\mathcal{G}_2,\,\text{lc}}=0 \\
    &W(\graph{3})=\frac{1}{M}\left(1-\frac{1}{M}\right)\,\qquad\, S(\graph{3}) = 1 \qquad\qquad \langle\sigma_1\sigma_2\rangle_c\Big|_{\mathcal{G}_3,\,\text{lc}}= \lambda^{L_0+L_1+L_2}\,. \nonumber
\end{align}
we obtain
\begin{align}\label{eq:2punti}
     \overline{\langle\sigma(x_1)\sigma(x_2)\rangle_c}&= \frac{(2D)^2}{M}\sum_{L}^{1,\dots,\infty}\nbp{L}{x_1}{x_2}\langle\sigma_1\sigma_2\rangle_c\Big|_{\mathcal{G}_1,\,\text{lc}}      + \nonumber\\
     &\hspace{-1cm}+\frac{(2D)^2}{2M^2}\frac{(2D)!}{(2D-4)!}\sum_{L_0,L_1,L_2}^{1,\dots,\infty}\sum_{x_0\in a\mathbb{Z}^D}\nbp{L_1}{x_1}{x_0}\nbp{L_0}{x_0}{x_0}\nbp{L_2}{x_0}{x_2}\langle\sigma_1\sigma_2\rangle_c\Big|_{\mathcal{G}_2,\,\text{lc}}+\nonumber\\
     &\hspace{-1cm}-\frac{(2D)^2}{M^2}\frac{(2D)!}{(2D-4)!}\sum_{L_0,L_1,L_2}^{1,\dots,\infty}\sum_{x_0\in  a\mathbb{Z}^D}\nbp{L_1}{x_1}{x_0}\nbp{L_0}{x_0}{x_0}\nbp{L_2}{x_0}{x_2}\langle\sigma_1\sigma_2\rangle_c\Big|_{\mathcal{G}_3,\,\text{lc}} +\mathcal{O}\left(\frac{1}{M^3}\right)  =\nonumber\\
     &= \frac{(2D)^2}{M} \sum_{L=1}^{\infty} \mathcal{N}_L(x_1,x_2) \, \lambda^{L} + \nonumber\\
    &\hspace{-1cm}- \frac{(2D)^2}{M^2}  \frac{(2D)!}{(2D-4)!}\sum_{x_0\in a\mathbb{Z}^D}\sum_{L_1=1}^{\infty} \mathcal{N}_{L_1}(x_1,x_0)\lambda^{L_1}\sum_{L_0=1}^{\infty}\mathcal{N}_{L_0}(x_0,x_0)\lambda^{L_0}\sum_{L_2=1}^{\infty}\mathcal{N}_{L_2}(x_0,x_2) \, \lambda^{L_2}+\mathcal{O}\left(\frac{1}{M^3}\right) \,, 
\end{align}

Introducing the generating function of the number of NBPs defined as \cite{Fitzner_2013}:
\begin{equation}
     B_{\lambda}(x_f,x_i)=\sum_{L=1}^{\infty}\mathcal{N}_L(x_f,x_i)\lambda^L\,,
\end{equation}
Eq. \eqref{eq:2punti} becomes:
\begin{equation}
    \overline{\langle\sigma(x_1)\sigma(x_2)\rangle_c} = \frac{(2D)^2}{M} B_{\lambda}(x_1,x_2) -\frac{(2D)^2}{M^2}  \frac{(2D)!}{(2D-4)!}\sum_{x_0\in a\mathbb{Z}^D}B_{\lambda}(x_1,x_0)B_{\lambda}(x_0,x_0)B_{\lambda}(x_0,x_2)+\mathcal{O}\left(\frac{1}{M^3}\right)\,.
\end{equation}
The next step is to Fourier transform the last expression.
The convention used to define the Fourier transform of a generic function $f(x)$ is:
\begin{equation}
    \hat{f}(k)=a^D\sum_{x\in a\mathbb{Z}^D}f(x)e^{ikx}\,,\qquad\quad f(x)=\int_{\left[-\frac{\pi}{a},\frac{\pi}{a}\right]}\frac{d^Dk}{(2\pi)^D}\hat{f}(k)e^{-ikx}\,;
\end{equation}
obtaining:
\begin{equation}\label{eq:2pointRAW}
    \overline{\langle\sigma(p)\sigma(q)\rangle_c} = (2\pi)^D\delta^D(p+q)\frac{(2D)^2}{M} \hat{B}_\lambda(p)\left(1-\frac{1}{M}\frac{(2D)!}{(2D-4)!} \frac{1}{a^D} \hat{B}_\lambda(q)\int_{[-\frac{\pi}{a},\frac{\pi}{a}]^D}\frac{d^D k}{(2\pi)^D} \hat{B}_\lambda(k) \right)  +\mathcal{O}\left(\frac{1}{M^3}\right)\,.
\end{equation}
Using the fact that the small momentum limit of the reciprocal of the generating function, in Fourier space, is
\begin{equation}
    \hat{B}^{-1}_{\lambda}\left(k\ll\frac{1}{a}\right) = \frac{1}{a^D} \frac{1-\lambda(2D-1)}{2D\lambda} + \frac{1}{a^D}\frac{\lambda+1}{4D^2\lambda(1-\lambda)}\,(ka)^2 + \mathcal{O}\big((ka)^4\big) \,,
\end{equation}
where the parameter $a$ is the lattice spacing, and using the definitions of $\rho$, $\tau$, $g$ and $\lambda_c$ from Eqs. \eqref{eq:DefRho}, \eqref{eq:DefTau}, \eqref{eq:Defg} and \eqref{eq:DefLambdac}, one finally recovers the critical behavior of the connected 2-point correlation function, as displayed in Eq. \eqref{eq:2point}. An analogous, more detailed computation can be found for the connected 4-point correlation function in App. \ref{app:4pointconnectedcorrelation}. 

\section{Conclusions}\label{sec:conclusion}
In the present paper, we applied the $M$-layer construction to the Ising model. Besides giving a methodological explanation on how to use this construction, we computed the 2 and 4-point correlation functions at leading order, including the one-loop corrections and we found the same results as the ones coming from field theory. 

In this work, we have paved the way for the application of the $M$-layer construction to systems less understood than the Ising model. Some previous works \cite{Angelini_2019,Angelini_2021,Rizzo_2019,Rizzo_2020,baroni_2023} have already obtained results from the use of $M$-layer, however, in these cases, the results did not need to take into account all constants in a systematic way as was done in this work. An example is the calculation of the upper critical dimension exploiting the Ginzburg criterion \cite{Angelini_2021}: for this calculation, it is sufficient to look at the dimension at which the one-loop correction to the 2-point correlation function diverges, thus the constants due to the vertices and the combinatorial factors are irrelevant. On the other hand, to obtain more accurate predictions on critical behavior, such as critical exponents, it is necessary to check all the various factors mentioned above before adding the contributions of the various diagrams, in order to identify the right coupling constant or the mass.
In this paper, we did it for the case of the ferromagnetic Ising model, for which we expect that an expansion around the Bethe lattice model, or around a fully connected model will give the same results because the ferromagnetic transitions have the same features on the two
MF models. We in fact verified this, validating the $M$-layer method. 
This is instructive and really important before proceeding to apply the same method to more complex systems, such as the spin-glass with an external field, for which we do not expect the two expansions to give the same results.

Moreover, in Eqs. (\ref{eq:DefRho},\ref{eq:DefTau},\ref{eq:Defg},\ref{eq:DefLambdac}) we can relate the ``field-theoretical'' constants of the model in the $M$-layer construction to some physical observables such as the lattice spacing $a$, the physical dimension $D$, and the eigenvalues and eigenvectors of the transfer matrix $T(\sigma,\tau)$, as defined in App. \ref{app:transfermatrix}. We thus can numerically control the present constants: this could be of crucial importance to understanding the finite-dimensional behavior of systems if the renormalization flux has more than one fixed point (FP) with a finite basin of attraction. Take as an example the case of spin-glass transition in a field: from standard field theory one finds that a Gaussian FP for the RG equations exists and is locally stable above $D_U=6$ \cite{bray1980renormalisation}. However, its basin of attraction shrinks to zero approaching $D_U$ from above \cite{moore2011disappearance}. Due to the finite basin of attraction of this FP, an additional, non-Gaussian FP could exist also above the upper critical dimension, as proposed by different authors \cite{angelini2015spin, charbonneau2019morphology, Angelini_2021}.  The FP governing the spin-glass behavior in finite dimensions in the presence of more than one stable FP depends on the initial values for $g$. In the $M$-layer approach, once RG equations are written, one can run them for $g$, knowing exactly the initial value and thus understanding which will be the final reached FP. One could also think to manually change the microscopic details of the model, changing $g_{\lambda}(\sigma)$, $Q(\sigma)$ and $P_B(\sigma)$, trying in this way to enter different basins of attraction in the parameters space. 

An important ingredient in this paper is the presence of the so-called \textit{boucle} diagrams. As explained in the main text, these contributions are corrections, needed to properly compute the average over the rewirings in the context of the $M$-layer construction. In future applications of this construction, it is important to include these contributions if even-degree vertices are computed, as in the case of the Ising model. In all previous applications of the $M$-layer method, odd-degree vertices were considered, for which \textit{boucle}-corrections to the vertices are not present. In odd-degree vertices theories there could still be boucle-corrections to the single line, e.g. diagram $\mathcal{G}_3$, but they should give a term that is subdominant with respect to one-loop cubic diagrams for the 2-points propagator in the large length limit.

Finally, let us discuss the role of the disorder in the $M$-layer approach. The rewirings are extracted at random and thus one could expect that this randomness should reflect in an additional parameter, w.r.t. the ones ($\rho$, $\tau$ and $g$) that are present in the standard non-random field theory. However, we showed that at one-loop order, this is not the case: no additional parameter associated with the disorder is generated.
This is, however, reasonable already thinking to the random regular graphs: we have seen that in the large $M$ limit, the $M$-layered graph tends to a Bethe lattice with fixed connectivity, that has a disordered topology. However, even if the Bethe lattice is disordered, for an Ising model the local cavity marginals, and the consequent ``local critical temperature'', are the same for all the spins. We thus expect that the effect of the disorder will appear only at the next orders in the $1/M$ expansion, leading to a disordered local critical temperature and a different universality class for the critical exponents. If instead the original Hamiltonian already contains disorder on the Bethe lattice (such as the Ising model with disordered couplings), the disorder associated with the rewirings should not change the universality class of the model.

\appendix

\section{$M$-layer construction}\label{app:mlayerconstruction}
In this Appendix, we present a detailed derivation of the results of the main text. In Section \ref{app:transfermatrix} we derive the expression of the joint probability distribution of two spins on a chain, Eq. \eqref{eq:jointprobdistribution}. To be precise we present the explicit derivation for the case of Ising-type variables, to generalize, in App.\ref{sec:SoftSpin}, to the case of soft spins, with a generic measure $d\mu(\sigma)$. In Section \ref{app:4pointconnectedcorrelation} we show how to compute the contributions of the other observable of interest, the 4-point connected correlation function on the diagrams of Fig. \ref{fig:4point}. Then we complete the evaluation plugging all the contributions together. Finally, in Section \ref{app:dimensionalanalysis} we perform the dimensional analysis, identifying the relevant constants to map the results of the $M$-layer construction to the standard results of the field theory associated with the Ising model.

\subsection{Transfer Matrix}\label{app:transfermatrix}
Here we want to derive the expression for the (unnormalized) joint probability distribution of two spin variables for the Ising model, in the absence of an external field on a Bethe lattice. We consider \textit{N} Ising variables $\sigma=\pm 1$, so the measure on each spin is 
\begin{equation}
    d\mu (\sigma) = \delta(\sigma^2-1)d\sigma\,;
\end{equation}
the Hamiltonian is given by Eq. \eqref{eq:ising_hamiltonian}. For definiteness, we consider a random regular graph (RRG) of connectivity $c>2$. The cavity equation reads:
\begin{align}
    Q(\sigma)&=\frac{1}{Z_Q}\sum_{\sigma_k=\pm 1}e^{\beta J \sigma \sigma_k }Q^{c-1}(\sigma_k)\label{eq:cavity}\\
    Z_Q&=\sum_{\sigma,\sigma_k=\pm 1}e^{\beta J \sigma \sigma_k }Q^{c-1}(\sigma_k)\,,
\end{align}
the function $Q(\sigma)$ is the cavity distribution on spin $\sigma$ where only one interaction is considered, such that the full marginal, which we call $P_{Bethe}(\sigma)$, is, up to a normalization constant:
\begin{equation}
     P_{Bethe}(\sigma)\propto Q(\sigma)^c\,.
\end{equation}
Since $Q(\sigma)$ is connected with the probability of the configuration of the spin $\sigma$, we expect it to be an even function for the zero external field case (in the paramagnetic phase), we can easily compute it:
\begin{equation}\label{eq:Q}
    Q(\sigma)=\frac{e^{\beta J \sigma}Q(1)^{c-1}+e^{-\beta J \sigma}Q(-1)^{c-1}}{\sum_{\sigma=\pm 1}\left(e^{\beta J \sigma}Q(1)^{c-1}+e^{-\beta J \sigma}Q(-1)^{c-1}\right)}=\frac{\cosh{(\beta J \sigma)}}{2\cosh{(\beta J )}}=\begin{cases}\frac{1}{2}\quad \text{ if } \sigma=1 \\
    \frac{1}{2} \quad\text{ if } \sigma=-1
    \end{cases}\,,
\end{equation}
where the second equality is justified using the fact that $Q(+1)=Q(-1)$. 

We define the transfer matrix as
\begin{equation}\label{eq:transfermatrix}
    T(\sigma,\tau)=\sqrt{Q(\sigma)^{c-2}}e^{\beta J \sigma \tau}\sqrt{Q(\tau)^{c-2}}=\frac{1}{2^{c-2}}\begin{pmatrix} e^{\beta J} & e^{-\beta J} \\
e^{-\beta J} & e^{\beta J} \end{pmatrix}\,
\end{equation}
with eigenvalues and corresponding normalized eigenvectors:
\begin{align}
    &\lambda_{max}=\frac{\cosh{(\beta J)}}{2^{c-3}}\qquad \longrightarrow \qquad \Vec{e}_{\lambda_{max}}=\frac{1}{\sqrt{2}}\begin{pmatrix}
        1\\
        1
    \end{pmatrix}\,;\\
    &\quad\lambda_{-}=\frac{\sinh{(\beta J)}}{2^{c-3}}\qquad \longrightarrow \qquad  \Vec{e}_{\lambda_{-}}=\frac{1}{\sqrt{2}}\begin{pmatrix}
        1\\
        -1
    \end{pmatrix}\,.
\end{align}
Using the decomposition of symmetric matrices we write the $L$-th power of $T(\sigma,\tau)$, scaled by $\lambda_{max}^L$:
\begin{equation}\label{eq:transfer_matrix_decomposition}
    \frac{T^L(\sigma,\tau)}{\lambda_{max}^L}= \Vec{e}_{\lambda_{max}}(\sigma)\Vec{e}_{\lambda_{max}}(\tau)+\lambda^L\, \Vec{e}_{\lambda_{-}}(\sigma)\Vec{e}_{\lambda_{-}}(\tau)\,,
\end{equation}
where we defined $\lambda\equiv\lambda_{-}/\lambda_{max}=\tanh{(\beta J)}$ . 

From \eqref{eq:cavity} and \eqref{eq:transfermatrix} we can conclude that $\Vec{e}_{\lambda_{max}}(\sigma)\propto \sqrt{Q(\sigma)^{c}} $:
\begin{equation}
    Q(\sigma)\propto\frac{1}{\sqrt{Q(\sigma)^{c-2}}}\sum_{\sigma_k=\pm1}T(\sigma,\sigma_k)\frac{1}{\sqrt{Q(\sigma_k)^{c-2}}}Q(\sigma_k)^{c-1}\,,
\end{equation}
that is:
\begin{equation}
    \sqrt{Q(\sigma)^c}\propto \sum_{\sigma_k=\pm1} T(\sigma,\sigma_k)\sqrt{Q(\sigma_k)^c}\,.
\end{equation}
Moreover, since $\Vec{e}_{\lambda_{max}}$ has norm 1:
\begin{equation}
    \Vec{e}_{\lambda_{max}}=\frac{\sqrt{Q(\sigma)^c}}{\sqrt{\sum_{\sigma'}Q(\sigma')^c}}\,.
\end{equation}

Next, we can compute the (scaled) partition function $Z_L(\sigma,\tau)$ for an open chain with fixed boundary spins $\sigma$ and $\tau$ (without entering cavity fields on $\sigma$ and $\tau$, except for the ones coming from the considered chain, this means that the connectivity of the central spins is $c$, while the one of $\sigma$ and $\tau$ is 1):
\begin{equation}\label{eq:partition}
    Z_L(\sigma,\tau)\equiv\frac{T^L(\sigma,\tau)/\lambda^L_{max}}{\sqrt{Q^{c-2}(\sigma)}\sqrt{Q^{c-2}(\tau)}}=\frac{\Vec{e}_{\lambda_{max}}(\sigma)\Vec{e}_{\lambda_{max}}(\tau)}{{\sqrt{Q^{c-2}(\sigma)}\sqrt{Q^{c-2}(\tau)}}}+\lambda^L\, \frac{\Vec{e}_{\lambda_{-}}(\sigma)\Vec{e}_{\lambda_{-}}(\tau)}{{\sqrt{Q^{c-2}(\sigma)}\sqrt{Q^{c-2}(\tau)}}}\,,
\end{equation}
that is
\begin{equation}
     Z_L(\sigma,\tau)=\frac{Q(\sigma)}{\sqrt{\sum_{\sigma'}Q(\sigma')^c}}\frac{Q(\tau)}{\sqrt{\sum_{\sigma'}Q(\sigma')^c}}+\lambda^L\, \frac{\Vec{e}_{\lambda_{-}}(\sigma)\Vec{e}_{\lambda_{-}}(\tau)}{{\sqrt{Q^{c-2}(\sigma)}\sqrt{Q^{c-2}(\tau)}}}\,.
\end{equation}

Notice that for $L\rightarrow\infty$ the marginal probability between the two spins is factorized because the second term of the RHS tends to zero, being $\lambda<1$. In this case, $Q(\sigma)$ must be the marginal probability distribution of the spin $\sigma$. This is indeed the case (in the analyzed chain the connectivity of the boundary spins is 1). 

We compactly rewrite the previous quantity for the case of the Ising model (without external field) :
\begin{equation}\label{eq:openchain}
    Z_L(\sigma,\tau)=P_B(\sigma)P_B(\tau)+\lambda^L\,{g_{\lambda}(\sigma)\,g_{\lambda}(\tau)}\,,
\end{equation}
where we defined (similarly to \cite{Altieri_2017}) $P_B(\sigma)$ and $g_{\lambda}(\sigma)$:
\begin{equation}\label{eq:Qg}
   P_B(\sigma)\equiv\frac{Q(\sigma)}{\sqrt{\sum_{\sigma'}Q(\sigma')^c}}\qquad\qquad\text{and}\qquad\qquad g_{\lambda}(\sigma)\equiv\frac{\Vec{e}_{\lambda_-}(\sigma)}{\sqrt{Q^{c-2}(\sigma)}}\,,
\end{equation}
From Eq. \eqref{eq:Q}, we have:
\begin{equation}\label{eq:PB_g_definitons_Ising}
    P_B(\sigma)=2^{\frac{c-3}{2}}\binom{1}{1}\qquad\text{and}\qquad g_{\lambda}(\sigma)=2^{\frac{c-3}{2}}\binom{1}{-1}\,.
\end{equation}
As a remark we notice that, while $P_B(\sigma)$ is even with respect its argument, $g_{\lambda}(\sigma)$ is odd. This is an important consequence of the $\mathbb{Z}_2$ symmetry of the Ising model. This symmetry reveals to be of fundamental importance in the computation of the observable, on a specific topology, using the cavity method.

The first important thing to check is the self-consistency of $Z_L(\sigma,\tau)$, namely if two chains of length $L_1$ and $L_2$ are joined together, once summed over the internal spin $\sigma_0$, the scaled partition function of the total chain of length $L_1+L_2$ should have the same form:
\begin{equation}
    \sum_{\sigma_0=\pm1} Z_{L_1}(\sigma_1,\sigma_0)\q{c-2}{0}Z_{L_2}(\sigma_0,\sigma_2)\stackrel{!}{=}Z_{L_1+L_2}(\sigma_1,\sigma_2)\,.
\end{equation}
This can be done plugging in $Z_{L_1}(\sigma_1,\sigma_0)$ and $Z_{L_2}(\sigma_0,\sigma_2)$ their definitions from Eq. \eqref{eq:partition} and summing over $\sigma_0$, to check that the definitions of $\pb{}{}$ and $g_{\lambda}(\sigma)$ are consistent:
\begin{align}
    &\sum_{\sigma_0=\pm1} Z_{L_1}(\sigma_1,\sigma_0)\q{c-2}{0}Z_{L_2}(\sigma_0,\sigma_2) =\nonumber \\
    &=\sum_{\sigma_0=\pm1} \left(P_B(\sigma_1)P_B(\sigma_0)+\lambda^{L_1}\,g_{\lambda}(\sigma_1)g_{\lambda}(\sigma_0)\right)Q(\sigma_0)^{c-2}\left(P_B(\sigma_0)P_B(\sigma_2)+\lambda^{L_2}\,g_{\lambda}(\sigma_0)g_{\lambda}(\sigma_2)\right)=\nonumber\\
    &= \left(\sum_{\sigma_0=\pm1} P_B(\sigma_0)^2\q{c-2}{0}\right)P_B(\sigma_1)P_B(\sigma_2)+\left(\sum_{\sigma_0=\pm1} g_{\lambda}(\sigma_0)^2Q(\sigma_0)^{c-2}\right)\lambda^{L_1+L_2}g_{\lambda}(\sigma_1)g_{\lambda}(\sigma_2)\,,
\end{align}
where the last equality is justified by the fact the function $g_{\lambda}(\sigma)$ is odd with respect to the argument. This implies two identities:
\begin{equation}\label{eq:selfconsistency}
    \sum_{\sigma=\pm1} P_B(\sigma)^2\q{c-2}{} = 1\qquad \text{and} \qquad \sum_{\sigma=\pm1} g_{\lambda}(\sigma)^2 Q(\sigma)^{c-2}=1\,,
\end{equation}
both can be easily verified using the definitions Eq. \eqref{eq:Qg}, remembering that $\sum_{\sigma}Q(\sigma)=1$ and $\sum_{\sigma}e^2_{\lambda_-}(\sigma)=1$.

Notice that the expression for the joint probability distribution, given in Eq. \eqref{eq:openchain}, can be generalized for the case of variables with a generic measure $d\mu(\sigma)$, taking into account the fact that the transfer matrix would have an infinite set of eigenvalues and eigenfunctions, so that
\begin{equation}
    Z_L(\sigma,\tau)=P_B(\sigma)P_B(\tau)+\sum_{\substack{i=2,\dots\\|\lambda_{i+1}|<|\lambda_i|<1}}\lambda_i^L\,{g_{\lambda_i}(\sigma)\,g_{\lambda_i}(\tau)}\,,
\end{equation}
where $\lambda_2$ corresponds to $\lambda$ and it is the one that assumes the critical value $\lambda_c=\frac{1}{2D-1}$ at the critical temperature $T_c$. In App. \ref{sec:SoftSpin} we explore the case of the soft spin variables, getting the same result as the Ising case.

\subsection{4-point connected correlation function}\label{app:4pointconnectedcorrelation}

\begin{figure}[H]
    \centering
    \includegraphics[scale=0.35]{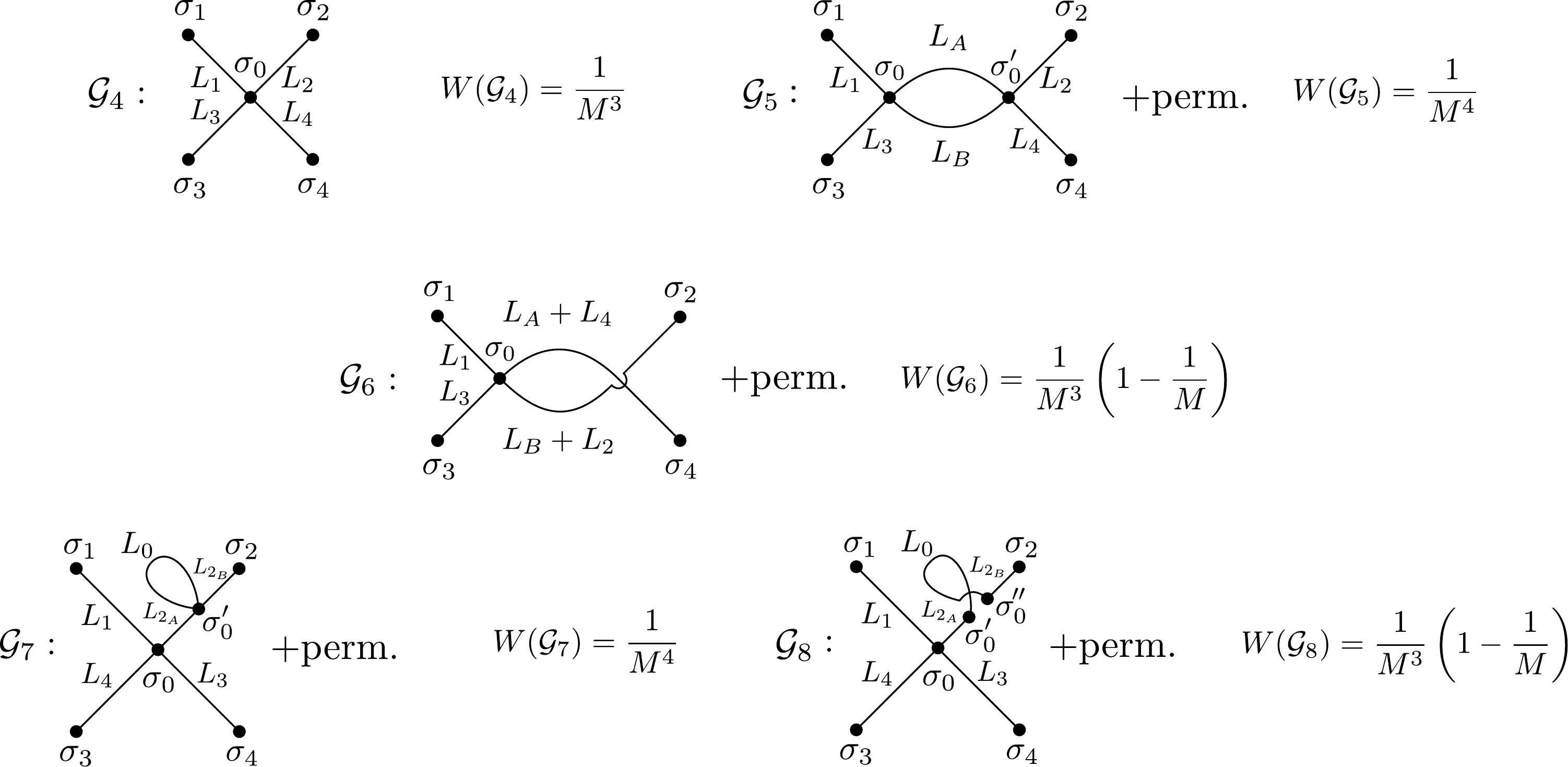}
    \caption{Topological diagram contributions to the 4-point connected correlation function. The order $\mathcal{O}(1/M^3)$ is the simple cross without loops connecting the four spins. At order $\mathcal{O}(1/M^4)$, that is at one loop, for each topology there are two distinct contributions: $\mathcal{G}_5$ and $\mathcal{G}_7$. As in the case of the 2-point correlation, one should also consider the so-called ``\textit{boucle}'' diagrams, $\mathcal{G}_6$ and $\mathcal{G}_8$, where the line doesn't close in a loop on the layered lattice but realizes it in the projection. A similar contribution for $\mathcal{G}_6$ is the one where the superimposition happens in the left internal vertex and it is not represented in the figure. Thus we should include a factor 2 for the $\mathcal{G}_6$ \textit{boucle} contribution of the 4-point function. The weight for \textit{boucle} diagrams is $W(\mathcal{G}_6)=W(\mathcal{G}_8)=1/M^3-1/M^4$.}
    \label{fig:4point}
\end{figure}

Following the method introduced in Sec. \ref{sec:mlayer}, we compute the 4-points connected correlation function. In this case the leading term is $\mathcal{O}\left(1/M^3\right)$ and we should include all contributions up to $\mathcal{O}\left(1/M^4\right)$ in order to consider the diagrams up to one loop level. The result is reported in Fig. \ref{fig:4point}, the corresponding weights are: $W(\mathcal{G}_4)=1/M^3$, $W(\mathcal{G}_5)=W(\mathcal{G}_7)=1/M^4$,  $W(\mathcal{G}_6)=W(\mathcal{G}_8)=1/M^3-1/M^4$.

Let us look now at the 4-point function, starting from diagram $\mathcal{G}_4$ in Fig. \ref{fig:4point}.

We have:
\begin{equation}
    \langle\sigma_1\sigma_2\sigma_3\sigma_4\rangle\Big|_{\mathcal{G}_4} = \frac{\lambda^{\sum_{i=1}^4 L_i}\left(\sum\limits_{\sigma=\pm1} \sigma\,g_{\lambda}(\sigma)\,Q(\sigma)^{c-1}\right)^4\,\left(\sum\limits_{\sigma_0=\pm1} \,g_\lambda(\sigma_0)^4\,Q(\sigma_0)^{c-4}\right)}{\left(\sum\limits_{\sigma=\pm1} \,P_B(\sigma)\,Q(\sigma)^{c-1}\right)^4\,\left(\sum\limits_{\sigma_0=\pm1} \,P_B(\sigma_0)^4\,Q(\sigma_0)^{c-4}\right)}=\lambda^{\sum_{i=1}^4 L_i}\,,
\end{equation}
where we used the definitions of $P_B(\sigma)$, $g_\lambda(\sigma)$ and $Q(\sigma)$, respectively Eqs. \eqref{eq:PB_g_definitons_Ising} and \eqref{eq:Q}. Now we compute the connected 4-point correlation function:
\begin{equation}
    \langle\sigma_1\sigma_2\sigma_3\sigma_4\rangle_c=\langle\sigma_1\sigma_2\sigma_3\sigma_4\rangle-\langle\sigma_1\sigma_2\rangle\langle\sigma_3\sigma_4\rangle-\langle\sigma_1\sigma_3\rangle\langle\sigma_2\sigma_4\rangle-\langle\sigma_1\sigma_4\rangle\langle\sigma_2\sigma_3\rangle\,,
\end{equation}
where the 3-point and 1-point connected correlation functions are zero for symmetry reasons, hence they are not included in the definition of the 4-point one. We compute the 2-point function between $\sigma_1$ and $\sigma_2$, the others are similar:
\begin{align}
    \langle\sigma_1\sigma_2\rangle\Big|_{\mathcal{G}_4} &=\lambda^{L_1+L_2}\frac{\Big(\sum\limits_{\sigma=\pm1} \sigma\,g_{\lambda}(\sigma)\,Q(\sigma)^{c-1}\Big)^2\,\Big(\sum\limits_{\sigma=\pm1} \,P_B(\sigma)\,Q(\sigma)^{c-1}\Big)^2\,\Big(\sum\limits_{\sigma_0=\pm1} \,g_{\lambda}(\sigma_0)^2\,P_B(\sigma_0)^2\,Q(\sigma_0)^{c-4}\Big)}{\Big(\sum\limits_{\sigma=\pm1} \,P_B(\sigma)\,Q(\sigma)^{c-1}\Big)^4\,\Big(\sum\limits_{\sigma_0=\pm1} \,P_B(\sigma_0)^4\,Q(\sigma_0)^{c-4}\Big)}= \nonumber \\
    &=\lambda^{L_1+L_2}\,.
\end{align}
We thus have:
\begin{align}
    \langle\sigma_1\sigma_2\sigma_3\sigma_4\rangle_c\Big|_{\mathcal{G}_4,\text{lc}} &=\langle\sigma_1\sigma_2\sigma_3\sigma_4\rangle_c\Big|_{\mathcal{G}_4} = \lambda^{\sum_{i=1}^4 L_i}\frac{\Big(\sum\limits_{\sigma=\pm1} \sigma\,g_{\lambda}(\sigma)\,Q(\sigma)^{c-1}\Big)^4}{\Big(\sum\limits_{\sigma=\pm1} P_B(\sigma)\,Q(\sigma)^{c-1}\Big)^4}\left( \frac{\sum\limits_{\sigma_0=\pm1} \,g_{\lambda}(\sigma_0)^4\,Q(\sigma_0)^{c-4}}{\sum\limits_{\sigma_0=\pm1} P_B(\sigma_0)^4\,Q(\sigma_0)^{c-4}} - 3 \right)= \nonumber \\
    &=-2\lambda^{\sum_{i=1}^4 L_i}\,,
\end{align}
$\langle\sigma_1\sigma_2\sigma_3\sigma_4\rangle_c\Big|_{\mathcal{G}_4}$ coincides with the corresponding definition of line-connected observable, since removing any of the lines of $\mathcal{G}_4$ the graph becomes disconnected, giving zero contribution.

\begin{figure}[H]
    \centering
    \includegraphics[scale=0.5]{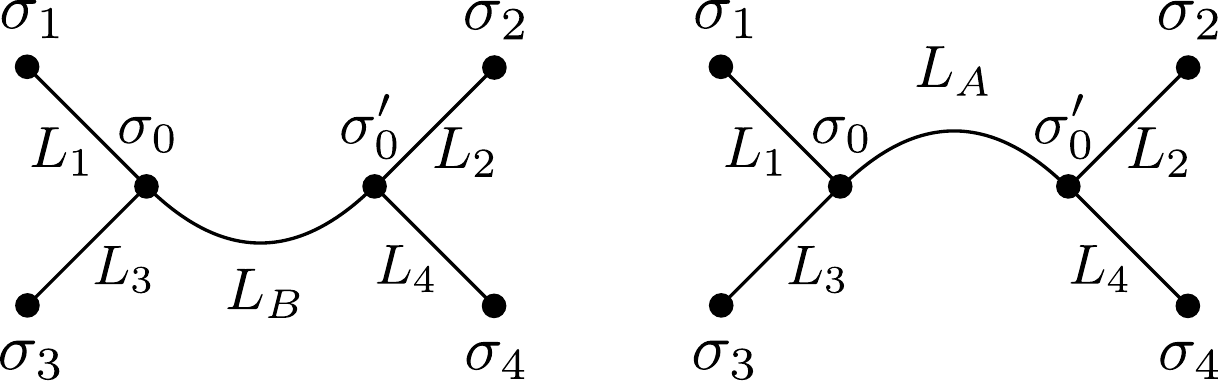}
    \caption{$\mathcal{G}_{5a}$ (left), $\mathcal{G}_{5b}$ (right) .}
    \label{fig:G5aG5b}
\end{figure}

At one loop level we compute the 4-point correlation defined on a graph with the topology, $\mathcal{G}_5$, depicted in Fig.\ref{fig:4point}. At this level we have to keep track of the definition of line-connected correlation and the definition of connected correlation, as explained for the 2-point correlation function. The former implies the subtraction of the connected 4-point correlation function computed on a graph with the two diagrams of Fig.\ref{fig:G5aG5b}. We can compute the total 4-point correlation function on $\mathcal{G}_5$ obtaining:
\begin{align}\label{eq:4PuntiLoopConnessa}
    &\langle\sigma_1\sigma_2\sigma_3\sigma_4\rangle\Big|_{\mathcal{G}_5} = \nonumber\\
    &\hspace{0.8cm}=\frac{\lambda^{\sum_{i=1}^4L_i}\,\Big(\sum\limits_{\sigma=\pm1} \sigma\,g_{\lambda}(\sigma)\,Q(\sigma)^{c-1}\Big)^4\,\left[\Big(\sum\limits_{\sigma=\pm1}\,g_{\lambda}(\sigma)^2 \,P_B(\sigma)^2\,Q(\sigma)^{c-4}\Big)^2+\lambda^{L_A+L_B}\Big(\sum\limits_{\sigma=\pm1}\,g_{\lambda}(\sigma)^4 \,Q(\sigma)^{c-4}\Big)^2\right]}{\Big(\sum\limits_{\sigma=\pm1} \,P_B(\sigma)\,Q(\sigma)^{c-1}\Big)^4\left[\Big(\sum\limits_{\sigma=\pm1}\,P_B(\sigma)^4 \,Q(\sigma)^{c-4}\Big)^2+\lambda^{L_A+L_B}\Big(\sum\limits_{\sigma=\pm1}\,g_{\lambda}(\sigma)^2 \,P_B(\sigma)^2\,Q(\sigma)^{c-4}\Big)^2\right]}  \nonumber\\
   &\hspace{0.8cm} =\lambda^{\sum_{i=1}^4L_i}\,;
\end{align}
\vspace{0.3cm}
while the 2-point functions can be of two different kinds:
\small
\begin{align}
    \langle\sigma_1\sigma_2\rangle\Big|_{\mathcal{G}_5} &= \frac{\lambda^{\sum_{i=1}^2L_i}\,\Big(\sum\limits_{\sigma=\pm1} \sigma\,g_{\lambda}(\sigma)\,Q(\sigma)^{c-1}\Big)^2\,\Big(\sum\limits_{\sigma=\pm1} \,P_B(\sigma)\,Q(\sigma)^{c-1}\Big)^2\,\Big(\sum\limits_{\sigma=\pm1}\,g_{\lambda}(\sigma)^2 \,P_B(\sigma)^2\,Q(\sigma)^{c-4}\Big)^2\,\left(\lambda^{L_A}\,+\lambda^{L_B}\right)}{\Big(\sum\limits_{\sigma=\pm1} \,P_B(\sigma)\,Q(\sigma)^{c-1}\Big)^4\left[\Big(\sum\limits_{\sigma=\pm1}\,P_B(\sigma)^4 \,Q(\sigma)^{c-4}\Big)^2+\lambda^{L_A+L_B}\Big(\sum\limits_{\sigma=\pm1}\,g_{\lambda}(\sigma)^2 \,P_B(\sigma)^2\,Q(\sigma)^{c-4}\Big)^2\right]}\nonumber \\
    & = \lambda^{\sum_{i=1}^2L_i}\,\frac{\,\lambda^{L_A}\,+\lambda^{L_B}}{1+\lambda^{L_A+L_B}}  \,,
\end{align}
\normalsize
and
\vspace{0.3cm}
\small
\begin{align}
    &\langle\sigma_1\sigma_3\rangle\Big|_{\mathcal{G}_5} = \nonumber\\
    &=\lambda^{L_1+L_3}\frac{\,\Big(\sum\limits_{\sigma=\pm1} \sigma\,g_{\lambda}(\sigma)\,Q(\sigma)^{c-1}\Big)^2\,\Big(\sum\limits_{\sigma=\pm1}\,g_{\lambda}(\sigma)^2 \,P_B(\sigma)^2\,Q(\sigma)^{c-4}\Big)\Big[\sum\limits_{\sigma=\pm1}\,P_B(\sigma)^4 \,Q(\sigma)^{c-4}+\lambda^{L_A+L_B}\,\sum\limits_{\sigma=\pm1}\,g_{\lambda}(\sigma)^4 \,Q(\sigma)^{c-4}\Big]}{\Big(\sum\limits_{\sigma=\pm1} \,P_B(\sigma)\,Q(\sigma)^{c-1}\Big)^4\left[\Big(\sum\limits_{\sigma=\pm1}\,P_B(\sigma)^4 \,Q(\sigma)^{c-4}\Big)^2+\lambda^{L_A+L_B}\Big(\sum\limits_{\sigma=\pm1}\,g_{\lambda}(\sigma)^2 \,P_B(\sigma)^2\,Q(\sigma)^{c-4}\Big)^2\right]} \nonumber \\
 &= \lambda^{L_1+L_3}\,,
\end{align}
\normalsize

All in all:
\begin{align}
   \langle\sigma_1\sigma_2\sigma_3\sigma_4\rangle_c\Big|_{\mathcal{G}_5} & =  \langle\sigma_1\sigma_2\sigma_3\sigma_4\rangle\Big|_{\mathcal{G}_5} - \langle\sigma_1\sigma_2\rangle\langle\sigma_3\sigma_4\rangle\Big|_{\mathcal{G}_5} -  \langle\sigma_1\sigma_3\rangle\langle\sigma_2\sigma_4\rangle\Big|_{\mathcal{G}_5} -\langle\sigma_1\sigma_4\rangle\langle\sigma_2\sigma_3\rangle\Big|_{\mathcal{G}_5} =\nonumber \\
   &= -2\frac{(\lambda^{L_A}+\lambda^{L_B})^2}{(1+\lambda^{L_A+L_B})^2}\,\lambda^{\sum_{i=1}^4 L_i}  \,.
\end{align}

Following the prescriptions of the $M$-layer construction we compute the line-connected 4-point connected correlation function subtracting the contributions of $\langle\sigma_1\sigma_2\sigma_3\sigma_4\rangle_c\Big|_{\mathcal{G}_5a}$ and $\langle\sigma_1\sigma_2\sigma_3\sigma_4\rangle_c\Big|_{\mathcal{G}_5b}$, as depicted in Fig. \ref{fig:G5aG5b}, to $\langle\sigma_1\sigma_2\sigma_3\sigma_4\rangle_c\Big|_{\mathcal{G}_5}$. For $\mathcal{G}_{5a}$ we find

\begin{align}
     \langle\sigma_1\sigma_2\sigma_3\sigma_4\rangle_c\Big|_{\mathcal{G}_{5a}} &= \frac{\lambda^{\sum_{i=1}^4L_i}\,\Big(\sum\limits_{\sigma=\pm1} \sigma\,g_{\lambda}(\sigma)\,Q(\sigma)^{c-1}\Big)^4\,\Big(\sum\limits_{\sigma=\pm1} \,g_{\lambda}(\sigma)^2\,P_B(\sigma)\,Q(\sigma)^{c-3}\Big)^2}{\Big(\sum\limits_{\sigma=\pm1} \,P_B(\sigma)\,Q(\sigma)^{c-1}\Big)^4\,\Big(\sum\limits_{\sigma=\pm1}\,P_B(\sigma)^3\,Q(\sigma)^{c-3}\Big)^2}\nonumber\\
     &\hspace{-1.2cm}-2\frac{\lambda^{2L_B+\sum_{i=1}^4L_i}\,\Big(\sum\limits_{\sigma=\pm1} \sigma\,g_{\lambda}(\sigma)\,Q(\sigma)^{c-1}\Big)^4\,\Big(\sum\limits_{\sigma=\pm1} \,P_B(\sigma)\,Q(\sigma)^{c-1}\Big)^4\,\Big(\sum\limits_{\sigma=\pm1} \,g_{\lambda}(\sigma)^2\,P_B(\sigma)\,Q(\sigma)^{c-3}\Big)^4}{\Big(\sum\limits_{\sigma=\pm1} \,P_B(\sigma)\,Q(\sigma)^{c-1}\Big)^8\,\Big(\sum\limits_{\sigma=\pm1}\,P_B(\sigma)^3\,Q(\sigma)^{c-3}\Big)^4}+\nonumber\\
     &\hspace{-1.2cm}-\frac{\lambda^{\sum_{i=1}^4L_i}\,\Big(\sum\limits_{\sigma=\pm1} \,P_B(\sigma)\,Q(\sigma)^{c-1}\Big)^4\,\Big(\sum\limits_{\sigma=\pm1} \,g_{\lambda}(\sigma)^2\,P_B(\sigma)\,Q(\sigma)^{c-3}\Big)^2\,\Big(\sum\limits_{\sigma=\pm1}\,P_B(\sigma)^3\,Q(\sigma)^{c-3}\Big)^2}{\Big(\sum\limits_{\sigma=\pm1} \,P_B(\sigma)\,Q(\sigma)^{c-1}\Big)^4\,\Big(\sum\limits_{\sigma=\pm1}\,P_B(\sigma)^3\,Q(\sigma)^{c-3}\Big)^4}=\nonumber\\
     &=-2\lambda^{2L_B+\sum_{i=1}^4L_i}\label{eq:G5a}\,,
\end{align}
analogously for $\mathcal{G}_{5b}$:

\begin{equation}
     \langle\sigma_1\sigma_2\sigma_3\sigma_4\rangle_c\Big|_{\mathcal{G}_{5b}} =  -2\lambda^{2L_A+\sum_{i=1}^4L_i} \,.
\end{equation}

At this point we have all the ingredients to compute the corresponding line-connected observable:

\begin{align}\label{eq:4PuntiLoopConnessaLineConnected}
    \langle\sigma_1\sigma_2\sigma_3\sigma_4\rangle_c\Big|_{\mathcal{G}_5,\,\text{lc}} &= \langle\sigma_1\sigma_2\sigma_3\sigma_4\rangle_c\Big|_{\mathcal{G}_5}-\langle\sigma_1\sigma_2\sigma_3\sigma_4\rangle_c\Big|_{\mathcal{G}_{5a}}-\langle\sigma_1\sigma_2\sigma_3\sigma_4\rangle_c\Big|_{\mathcal{G}_{5b}} \nonumber\\
    &=\lambda^{\sum_{i=1}^4 L_i} \Bigg[-2\frac{(\lambda^{L_A}+\lambda^{L_B})^2}{(1+\lambda^{L_A+L_B})^2}+2\lambda^{2L_A}+2\lambda^{2L_B}\Bigg] \,,
\end{align}
which can be Taylor-expanded up to $\mathcal{O}(\lambda^{6L})$, in the limit of $L_A\sim L_B\sim L_i\sim L\rightarrow\infty$:
\begin{align}
    \langle\sigma_1\sigma_2\sigma_3\sigma_4\rangle_c\Big|_{\mathcal{G}_5,\,\text{lc}}= -4\lambda^{L_A+L_B+\sum_{i=1}^4L_i}+\mathcal{O}(\lambda^{8L}) \,.
\end{align}

Before computing the contributions of the \textit{boucle} diagrams for the 4-point function let us notice that two more diagrams, with the same topology of $\mathcal{G}_5$ as explicitly shown in Fig. \ref{fig:4point}, contribute. One is obtained by exchanging $x_1$ and $x_2$ and, similarly, the second is obtained by exchanging $x_1$ and $x_4$. These two additional cases are not included in the symmetry factor due to the different entrance directions of the lines in the internal vertices, thus they must be added by hand. Notice that these contributions are present in the continuum field theory associated with the Ising model due to Wick's theorem. The same symmetry reasoning is true for the next contributions, the \textit{boucle} diagrams $\mathcal{G}_6$. By writing $\mathcal{G}_5$ and $\mathcal{G}_6$ we include, in the following, also their external vertices permutations.
The \textit{boucle} diagrams for the 4-point function, reported in Fig. \ref{fig:4point}, contribute with:

\begin{align}\label{eq:4PuntiCorrezioni}
    &\langle\sigma_1\sigma_2\sigma_3\sigma_4\rangle_c\Big|_{\mathcal{G}_6,\text{lc}} = \nonumber\\
    &\hspace{1cm}=\frac{\lambda^{L_A+L_B+\sum_{i=1}^4 L_i}\,\Big(\sum\limits_{\sigma=\pm1} \sigma\,g_{\lambda}(\sigma)\,Q(\sigma)^{c-1}\Big)^4}{\Big(\sum\limits_{\sigma=\pm1} \,P_B(\sigma)\,Q(\sigma)^{c-1}\Big)^4}\left( \frac{\sum\limits_{\sigma=\pm1}\,g_{\lambda}(\sigma)^4 \,Q(\sigma)^{c-4}}{\sum\limits_{\sigma=\pm1}\,P_B(\sigma)^4 \,Q(\sigma)^{c-4}} - 3\frac{\Big(\sum\limits_{\sigma=\pm1}\,g_{\lambda}(\sigma)^2 \,P_B(\sigma)^2\,Q(\sigma)^{c-4}\Big)^2}{\Big(\sum\limits_{\sigma=\pm1}\,P_B(\sigma)^4 \,Q(\sigma)^{c-4}\Big)^2} \right)\nonumber \\
    &\hspace{1cm} = -2\lambda^{L_A+L_B+\sum_{i=1}^4 L_i} \,.
\end{align}

Similarly, there is also the contribution due to the diagram where the overlap of the lines is in $\sigma_0$. Since in this case the value of the observable is exactly the same as defined on diagram $\graph{6}$ we will simply add a factor 2 in front of the contribution of $\graph{6}$ (see Eq. \eqref{eq:4punti}).

Now we compute the line-connected 4-point connected correlation function on the four possible diagrams of the kind of $\mathcal{G}_7$, that is where one of the external lines includes a loop of the same kind of $\mathcal{G}_2$, as depicted in Fig. \ref{fig:4point}.

\begin{align}\label{eq:4PuntiExtLegDress}
    &\langle\sigma_1\sigma_2\sigma_3\sigma_4\rangle_c\Big|_{\mathcal{G}_7,\,\text{lc}} = \nonumber\\
    &\hspace{1cm}=\lambda^{\sum_{i=1}^4 L_i} \frac{\Big(\sum\limits_{\sigma=\pm1} \sigma\,g_{\lambda}(\sigma)\,Q(\sigma)^{c-1}\Big)^4}{\Big(\sum\limits_{\sigma=\pm1} \,P_B(\sigma)\,Q(\sigma)^{c-1}\Big)^4} \left( 
 \frac{\sum\limits_{\sigma=\pm1}\,g_{\lambda}(\sigma)^4 \,Q(\sigma)^{c-4}}{\sum\limits_{\sigma=\pm1}\,P_B(\sigma)^4 \,Q(\sigma)^{c-4}} - 3\right)  \lambda^{L_0} \left(
\frac{\sum\limits_{\sigma=\pm1}\,g_{\lambda}(\sigma)^4 \,Q(\sigma)^{c-4}}{\sum\limits_{\sigma=\pm1}\,P_B(\sigma)^4 \,Q(\sigma)^{c-4}} -1  \right) = 0 \,.
\end{align}

The line connected 4-point connected correlation function on the four possible diagrams of the kind $\mathcal{G}_8$, that is where one of the external lines includes a \textit{boucle}, is:

\begin{equation}\label{eq:4PuntiExtLegDressBoucle}
    \langle\sigma_1\sigma_2\sigma_3\sigma_4\rangle_c\Big|_{\mathcal{G}_8,\,\text{lc}} = \lambda^{L_0+\sum_{i=1}^4 L_i} \frac{\Big(\sum\limits_{\sigma=\pm1} \sigma\,g_{\lambda}(\sigma)\,Q(\sigma)^{c-1}\Big)^4}{\Big(\sum\limits_{\sigma=\pm1} \,P_B(\sigma)\,Q(\sigma)^{c-1}\Big)^4} \left( 
 \frac{\sum\limits_{\sigma=\pm1}\,g_{\lambda}(\sigma)^4 \,Q(\sigma)^{c-4}}{\sum\limits_{\sigma=\pm1}\,P_B(\sigma)^4 \,Q(\sigma)^{c-4}} - 3\right) = -2\lambda^{L_0+\sum_{i=1}^4 L_i}\,.
\end{equation}
For $\mathcal{G}_7$ and $\mathcal{G}_8$ all four possible permutations of the external vertices should be included, considering, in this way, the four possible positions of the loop on the external lines. 

Before going on, we want to comment on the role of $\mathcal{G}_{5a}$ and $\mathcal{G}_{5b}$.
We computed the contribution of the 4-point correlation on them because it was needed for the computation of the line-connected observable on $\mathcal{G}_{5}$, see Eq. \eqref{eq:4PuntiLoopConnessaLineConnected}, but the attentive reader could ask why we did not consider them giving a contribution also in the 0-loop order for the 4-point observable. In fact, such a diagram would contribute with the same weight as $\mathcal{G}_4$, that is $1/M^3$, as can be checked with the same arguments given for computing the weight of the other diagrams in the main text. The difference, with respect to $\langle\sigma_1\sigma_2\sigma_3\sigma_4\rangle_c\Big|_{\mathcal{G}_4,\text{lc}}$ is that the potential contribution of $\langle\sigma_1\sigma_2\sigma_3\sigma_4\rangle_c\Big|_{\mathcal{G}_{5a},\text{lc}}=\langle\sigma_1\sigma_2\sigma_3\sigma_4\rangle_c\Big|_{\mathcal{G}_{5a}}$ in Eq. \eqref{eq:G5a}, would go as $\mathcal{O}(\lambda^{6L})$, in the limit of $L_A\sim L_B\sim L_i\sim L\rightarrow\infty$, while $\langle\sigma_1\sigma_2\sigma_3\sigma_4\rangle_c\Big|_{\mathcal{G}_{4},\text{lc}}=\mathcal{O}(\lambda^{4L})$.
$\mathcal{G}_{5a}$ can thus be neglected in the large-lengths limit w.r.t. $\mathcal{G}_{4}$. Let us underline that $\mathcal{G}_{5a}$ is a diagram with cubic vertices: the fact that it is subdominant with respect to the diagram $\mathcal{G}_{4}$ is in agreement with the fact that we expect the $M$-layer construction to give the same results as a standard field theoretical approach, for which cubic diagrams are absent.

At this point, we can put all the contributions together, including the weights, in terms of inverse power of \textit{M}, $W(\mathcal{G})$, of each diagram $\mathcal{G}$ and the corresponding symmetry factor $S(\mathcal{G})$. Remembering that
\begin{align}
    &W(\graph{4})=\frac{1}{M^3}\,\qquad \qquad \qquad\quad S(\graph{4}) = 1 \quad\qquad \langle\sigma_1\sigma_2\sigma_3\sigma_4\rangle_c\Big|_{\mathcal{G}_4,\,\text{lc}}= -2\lambda^{\sum_{i=1}^4L_i}\,; \nonumber\\
    &W(\graph{5})=\frac{1}{M^4}\,\qquad \qquad \qquad \quad S(\graph{5}) = 2 \quad\qquad \langle\sigma_1\sigma_2\sigma_3\sigma_4\rangle_c\Big|_{\mathcal{G}_5,\,\text{lc}}\simeq -4 \lambda^{L_A+L_B+\sum_{i=1}^4L_i}\,; \nonumber\\
    &W(\graph{6})=\frac{1}{M^3}\left(1-\frac{1}{M}\right)\,\quad\quad S(\graph{6}) = 1 \quad\qquad \langle\sigma_1\sigma_2\sigma_3\sigma_4\rangle_c\Big|_{\mathcal{G}_6,\,\text{lc}}=-2 \lambda^{L_A+L_B+\sum_{i=1}^4L_i}\,; \nonumber\\
    &W(\graph{7})=\frac{1}{M^4}\,\qquad  \qquad \qquad  \quad S(\graph{7}) = 2 \quad\qquad \langle\sigma_1\sigma_2\sigma_3\sigma_4\rangle_c\Big|_{\mathcal{G}_7,\,\text{lc}}=0\,; \nonumber\\
    &W(\graph{8})=\frac{1}{M^3}\left(1-\frac{1}{M}\right)\,\quad\quad S(\graph{8}) = 1 \quad\qquad \langle\sigma_1\sigma_2\sigma_3\sigma_4\rangle_c\Big|_{\mathcal{G}_8,\,\text{lc}}=-2\lambda^{L_0+\sum_{i=1}^4 L_i} \,,
\end{align}
the 4-point function is:
\begin{align}\label{eq:4punti}
    \overline{ \langle\sigma(x_1)\sigma(x_2)\sigma(x_3)\sigma(x_4)\rangle_c}&= \frac{(2D)^4}{M^3}\frac{(2D)!}{(2D-4)!}\sum_{L_1,L_2,L_3,L_4}^{1,\dots,\infty}\sum_{x_0\in  a\mathbb{Z}^D}\prod_{i=1}^4\nbp{L_i}{x_i}{x_0}\langle\sigma_1\sigma_2\sigma_3\sigma_4\rangle_c\Big|_{\mathcal{G}_4,\,\text{lc}}      + \nonumber\\
     &\hspace{-5cm}\qquad\qquad+\frac{(2D)^4}{2M^4}\left(\frac{(2D)!}{(2D-4)!}\right)^2\sum_{\Vec{L}}\sum_{x_0,x_0'\in  a\mathbb{Z}^D}\prod_{i=1,3}\nbp{L_i}{x_i}{x_0}\prod_{i=2,4}\nbp{L_i}{x_i}{x_0'}\nbp{L_A}{x_0}{x_0'}\nbp{L_B}{x_0}{x_0'}\langle\sigma_1\sigma_2\sigma_3\sigma_4\rangle_c\Big|_{\mathcal{G}_5,\,\text{lc}}    + \nonumber\\
     &\hspace{-5cm}\qquad\qquad-2\frac{(2D)^4}{M^4}\left(\frac{(2D)!}{(2D-4)!}\right)^2\sum_{\Vec{L}}\sum_{x_0,x_0'\in  a\mathbb{Z}^D}\prod_{i=1,3}\nbp{L_i}{x_i}{x_0}\prod_{i=2,4}\nbp{L_i}{x_i}{x_0'}\nbp{L_A}{x_0}{x_0'}\nbp{L_B}{x_0}{x_0'}\langle\sigma_1\sigma_2\sigma_3\sigma_4\rangle_c\Big|_{\mathcal{G}_6,\,\text{lc}}+ \nonumber\\
     &\hspace{-5cm}\qquad\qquad+\frac{(2D)^4}{2M^4}\left(\frac{(2D)!}{(2D-4)!}\right)^2\sum_{\Vec{L'}}\sum_{x_0,x_0'\in  a\mathbb{Z}^D}\prod_{i=1,3,4}\nbp{L_i}{x_i}{x_0}\nbp{L_{2_A}}{x_0'}{x_0}\nbp{L_0}{x_0'}{x_0'}\nbp{L_{2_{B}}}{x_2}{x_0'}\langle\sigma_1\sigma_2\sigma_3\sigma_4\rangle_c\Big|_{\mathcal{G}_7,\,\text{lc}}    + \nonumber\\
     &\hspace{-5cm}\qquad\qquad-\frac{(2D)^4}{M^4}\left(\frac{(2D)!}{(2D-4)!}\right)^2\sum_{\Vec{L'}}\sum_{x_0,x_0'\in  a\mathbb{Z}^D}\prod_{i=1,3,4}\nbp{L_i}{x_i}{x_0}\nbp{L_{2_A}}{x_0'}{x_0}\nbp{L_0}{x_0'}{x_0'}\nbp{L_{2_B}}{x_2}{x_0'}\langle\sigma_1\sigma_2\sigma_3\sigma_4\rangle_c\Big|_{\mathcal{G}_8,\,\text{lc}} \nonumber\\
     &\hspace{-5cm}\qquad\qquad + \text{permutations of $\mathcal{G}_5$, $\mathcal{G}_6$, $\mathcal{G}_7$ and $\mathcal{G}_8$} +\mathcal{O}\left(\frac{1}{M^5}\right)\,,
\end{align}
where $\Vec{L}=\{L_1,L_2,L_3,L_4,L_A,L_B\}$, $\Vec{L'}=\{L_1,L_{2_A},L_{2_B},L_3,L_4,L_0\}$. 
We can also write the 4-point function in Fourier space. The convention used here for the Fourier transform is the following:
\begin{equation}
    \hat{f}(k)=a^D\sum_{x\in a\mathbb{Z}^D}f(x)e^{ikx}\,,\qquad\quad f(x)=\int_{\left[-\frac{\pi}{a},\frac{\pi}{a}\right]}\frac{d^Dk}{(2\pi)^D}\hat{f}(k)e^{-ikx}\,;
\end{equation}
which implies 
\begin{equation}
    \left(\frac{2\pi}{a}\right)^D \,\delta^D(k)=\sum_{x\in a\mathbb{Z}^D}e^{ikx}\,.
\end{equation}
Moreover, we can use the definition of the generating function of the NBP \cite{Altieri_2017,Fitzner_2013}:
\begin{equation}\label{eq:generatingNBP}
    B_{\lambda}(x_f,x_i)=\sum_{L=1}^{\infty}\mathcal{N}_L(x_f,x_i)\lambda^L\,.
\end{equation}
Going into Fourier space, the 4-point function reads:
\small
\begin{align}
     &\overline{\langle\sigma(k_1)\sigma(k_2)\sigma(k_3)\sigma(k_4)\rangle_c}= \nonumber\\
     &\hspace{0.5cm}- \frac{2(2D)^4}{M^3}\frac{(2D)!}{(2D-4)!}\, \left(\frac{2\pi}{a}\right)^D\delta^D\left(\sum_{i=1}^4\,k_i\right) \prod_{i=1}^{4}\hat{B}_{\lambda}(k_i) \Bigg( 1 - \frac{1}{2} \frac{2}{M}  \frac{(2D)!}{(2D-4)!} \frac{1}{a^D} \sum_{j=2}^4\int \frac{d^Dq}{(2\pi)^D} \hat{B}_{\lambda}(q) \hat{B}_{\lambda}(q+k_1+k_j)\Bigg)+\nonumber\\
     &\hspace{0.5cm}+ \sum_{j=1}^4 \frac{2(2D)^4}{M^4}\left(\frac{(2D)!}{(2D-4)!}\right)^2 \frac{(2\pi)^D}{(a^D)^2}\delta^D\left(\sum_{i=1}^4\,k_i\right) \prod_{i=1}^4\hat{B}(k_i)\hat{B}(k_j)\int\frac{d^Dq}{(2\pi)^D}\hat{B}(q)+\mathcal{O}\left(\frac{1}{M^5}\right)\,,
\end{align}
\normalsize
Notice that with the sum over $j$ we include all the permutations of the external vertices on $\mathcal{G}_5$, $\mathcal{G}_6$, $\mathcal{G}_7$ and $\mathcal{G}_8$. Defining
\begin{equation}\label{eq:definizione_g}
    \tilde{g}\equiv \frac{2}{M}\frac{(2D)!}{(2D-4)!} 
\end{equation}
we get:
\begin{align}\label{eq:4}
     &\overline{\langle\sigma(k_1)\sigma(k_2)\sigma(k_3)\sigma(k_4)\rangle_c}= \nonumber\\
     &\hspace{1.5cm}\frac{(2D)^4}{M^2}\,\left(\frac{2\pi}{a}\right)^D\delta^D\left(\sum_{i=1}^4\,k_i\right) \prod_{i=1}^{4}\hat{B}_{\lambda}(k_i)\Bigg( -\tilde{g} + \frac{1}{2} \tilde{g}^2 \frac{1}{a^D} \sum_{j=2}^4 \int \frac{d^Dq}{(2\pi)^D} \hat{B}_{\lambda}(q) \hat{B}_{\lambda}(q+k_1+k_j)\Bigg)+\nonumber\\
     &\hspace{1.5cm}+ \sum_{j=1}^4 \frac{1}{2}\frac{(2D)^4}{M^2}\tilde{g}^2\frac{(2\pi)^D}{(a^D)^2}\delta^D\left(\sum_{i=1}^4\,k_i\right) \prod_{i=1}^4\hat{B}(k_i)\hat{B}(k_j)\int\frac{d^Dq}{(2\pi)^D}\hat{B}(q)+\mathcal{O}\left(\frac{1}{M^5}\right)\,.
\end{align}

This is the result for the computation of the 4-point connected correlation function for the $D$ dimensional Ising model due to the $M$-layer construction, up to order $\mathcal{O}(1/M^5)$ for $M\rightarrow\infty$. To be precise we also made use of the fact that, since we are interested in the critical behavior, we can perform the large-lengths limit.
In the next section, exploiting the critical behavior, we consider the small momentum limit of the generating functions of the NBP and we perform the dimensional analysis to complete the mapping to the continuum field theory.

\subsection{Dimensional analysis}\label{app:dimensionalanalysis}

The Fourier transform of the generating function of the number of NBPs can be computed \cite{Altieri_2017,Fitzner_2013}:
\begin{equation}\label{eq:NBWdef}
     \hat{B}_{\lambda}(k)=a^D\sum_{x\in a\mathbb{Z}^D}\sum_L^{1,\dots,\infty}\mathcal{N}_L(x,0)\lambda^L\,e^{i kx}=a^D\frac{2D\lambda\left(F(k)-\lambda\right)}{1+\lambda^2(2D-1)-2D\lambda F(k)}\,,
\end{equation}
where:
\begin{equation}
    F(k)=\frac{1}{D}\sum_{\mu=1}^D\cos{(k_{\mu}a)}\,,
\end{equation}
given $a$, the lattice spacing. At this point one can argue that the interesting limit for the critical behavior is the one for $k\rightarrow 0$, for which, using \eqref{eq:NBWdef}:
\begin{equation}
    \hat{B}^{-1}_{\lambda}\left(k\ll\frac{1}{a}\right) = \frac{1}{a^D} \frac{1-\lambda(2D-1)}{2D\lambda} + \frac{1}{a^D}\frac{\lambda+1}{4D^2\lambda(1-\lambda)}\,(ka)^2 + \mathcal{O}\big((ka)^4\big) \,,
\end{equation}
to reproduce the Gaussian propagator we can define:
\begin{equation}
   \tilde{\tau}\equiv \frac{1}{a^D} \frac{1-\lambda(2D-1) }{2D\lambda}\,\qquad\text{and}\qquad \tilde{\rho}\equiv \frac{1}{a^{D-2}}\frac{\lambda+1}{4D^2\lambda(1-\lambda)}\,,
\end{equation}
so that:
\begin{equation}
    \hat{B}^{-1}_{\lambda}\left(k\ll\frac{1}{a}\right) = \tilde{\rho} k^2+\tilde{\tau} + \mathcal{O}\big((ka)^4\big) \,.
\end{equation}

Let us consider the 2-point function for simplicity. Plugging $\hat{B}^{-1}_{\lambda}\left(k\ll\frac{1}{a}\right)$, neglecting $ \mathcal{O}\big((ka)^4\big) $, inside Eq. \eqref{eq:2pointRAW} we get:
\begin{equation}
    \overline{\langle\sigma(p)\sigma(q)\rangle_c} = (2\pi)^D\delta^D(p+q)\frac{(2D)^2}{M} \frac{1}{\tilde{\rho} p^2+\tilde{\tau}}\left(1-\frac{1}{2}\tilde{g} \frac{1}{a^D} \frac{1}{\tilde{\rho} q^2+\tilde{\tau}}\int_{[-\frac{\pi}{a},\frac{\pi}{a}]^D}\frac{d^D k}{(2\pi)^D} \frac{1}{\tilde{\rho} k^2+\tilde{\tau}} \right)  +\mathcal{O}\left(\frac{1}{M^3}\right)\,,
\end{equation}
from which we see how to redefine $\tilde{\tau}$, $\tilde{\rho}$ and $\tilde{g}$ in order to reproduce the continuum field theory result, displayed in Eq. \eqref{eq:2point}:
\begin{equation}
    \tau\equiv \frac{M}{(2D)^2}\tilde{\tau}\,,\qquad\qquad \rho\equiv \frac{M}{(2D)^2}\tilde{\rho}\qquad\text{and}\qquad g\equiv \frac{1}{a^D}\frac{M^2}{(2D)^4}\tilde{g}\,.
\end{equation}

 Now we want to express the dimensions of these three parameters in terms of length unit, $[L]$, and field unit, $[\sigma]$. To this aim, we should remember that, when computing the observables using the RS cavity method, we simplified all the terms containing sums over $\sigma$ of powers of $P_B(\sigma)$, $g_{\lambda}(\sigma)$ and $Q(\sigma)$, making use of their definitions. One of these terms is $\sum_{\sigma=\pm1}\sigma g_{\lambda}(\sigma)Q(\sigma)^{c-1}$, which had dimensions $[\sigma]$. Notice that this term, appearing with power 2 at the numerator of $\overline{\langle\sigma_1\sigma_2\rangle_c}$ (as can be checked by looking at Eq. \eqref{eq:g1lc}, or Eq. \eqref{eq:g2NONlc} and so on), is the one that gives the 2-point function the right dimensions, that is $[\sigma]^2$. Keeping track of these dimensional terms, inside the redefined parameter, we can understand that:
\begin{align}
    & [\,\tau\,]=[\,L\,]^{-D}[\,\sigma\,]^{-2}\\[5pt]
    & [\,\rho\,]=[\,L\,]^{2-D}[\,\sigma\,]^{-2}\\[5pt]
    & [\,g\,]=[\,L\,]^{-D}[\,\sigma\,]^{-4}\,.
\end{align}
The presence of these terms with the same dimensions as the field, $[\,\sigma\,]$, is more explicit for soft spin variables, see Eqs. \eqref{eq:softspincoupling} and \eqref{eq:softspinconstants} in App. \ref{sec:SoftSpin}.
The same can be done for the 4-point function, repeating the same steps, starting from \eqref{eq:4punti} and performing the $k\rightarrow0$ limit, with the same definitions of $\tau$, $\rho$ and $g$:
\begin{align}
    &\overline{\langle\sigma(k_1)\sigma(k_2)\sigma(k_3)\sigma(k_4)\rangle_c} = \nonumber\\
    &\hspace{1.5cm}=\frac{1}{M^2}\,(2\pi)^D\,\delta^D\left(\sum_{i=1}^4 k_i\right)\,\prod_{i=1}^4 \,G(k_i)\left[ -g +\frac{1}{2}g^2 \Big(I(k_1+k_2)+ I(k_1+k_3)+ I(k_1+k_4)\Big) \right]  +\mathcal{O}\left(\frac{1}{M^5}\right) \,,
\end{align}
where
\begin{align}
    I(q)&\equiv \int_{[-\frac{\pi}{a},\frac{\pi}{a}]^D} \frac{d^Dp}{(2\pi)^D}\frac{1}{\rho p^2+\tau}\cdot\frac{1}{\rho(p+q)^2+\tau}\,,\\
    G(q)&\equiv \frac{1}{\rho q^2+\tau} \Bigg(1  -\frac{1}{2} g \frac{1}{\rho q^2+\tau}\int_{[-\frac{\pi}{a},\frac{\pi}{a}]^D} \frac{d^Dk}{(2\pi)^D}\frac{1}{\rho k^2+\tau} \Bigg)\,.
\end{align}

To complete the mapping to the standard theory we recall some ideas. Let us consider the $g\phi^4$ action:
\begin{equation}
    \mathcal{L}=\int d^Dx \,\,\left(\frac{1}{2}\,\gamma\, \nabla \phi(x)\cdot\nabla \phi(x) + \frac{1}{2}\,m^2\,\phi^2(x) + \frac{1}{4!}\,g_0\,\phi^4(x)\right)\,,
\end{equation}
from which the Gaussian propagator reads:
\begin{equation}
    \hat{G}_0(k)=\frac{1}{\gamma\,k^2+m^2}\,,
\end{equation}
where $\gamma$, $m^2$ and $g_0$ depend on the temperature of the system, while $\phi(x)$ is the value of the field at the point $x$ of the $D$-dimensional space where it is defined. Their dimensions can be easily expressed in terms of length unit $[L]$, using that $\mathcal{L}$ is dimensionless:
\begin{align}
    &1\stackrel{!}{=}\left[ \int d^Dx \,\gamma (\nabla \phi)^2 \right]=[\,L\,]^{D-2}[\,\gamma\,][\,\phi\,]^2\quad\rightarrow\quad [\,\gamma\,]=[\,L\,]^{2-D}[\,\phi\,]^{-2}\\
    &1\stackrel{!}{=}\left[ \int d^Dx\, m^2 \, \phi^2 \right]=[\,L\,]^{D}[\,m^2\,][\,\phi\,]^2\quad\rightarrow\quad [\,m^2\,]=[\,L\,]^{-D}[\,\phi\,]^{-2}\\
    &1\stackrel{!}{=}\left[ \int d^Dx \,g_0 \phi^4 \right]=[\,L\,]^{D}[\,\phi\,]^4\quad\rightarrow\quad [\,g_0\,]=[\,L\,]^{-D}[\,\phi\,]^{-4}\,.
\end{align}

One can notice that the constant $\rho$, defined in the $M$-layer framework, corresponds to the temperature-dependent parameter $\gamma$ of the continuum $g_0\phi^4$ field theory, as well as $\tau$ corresponds to $m^2$ and $g$ to $g_0$, once the spin field $\sigma$ is identified with the field $\phi$. This last remark completes the mapping between the $M$-layer framework and the continuum field theory.

\section{Caveats}\label{app:caveats}

This section is devoted to underline two subtleties about the whole approach. The first is a non-trivial caveat about the geometry of the $M$-layer construction; the second is a relevant generalization of the previous arguments to soft spin variables, introducing features that are absent for the Ising model.

\subsection{Degree of the interaction in the \textit{M}-layer framework}
\begin{figure}[H]
    \centering
    \includegraphics[scale=0.45]{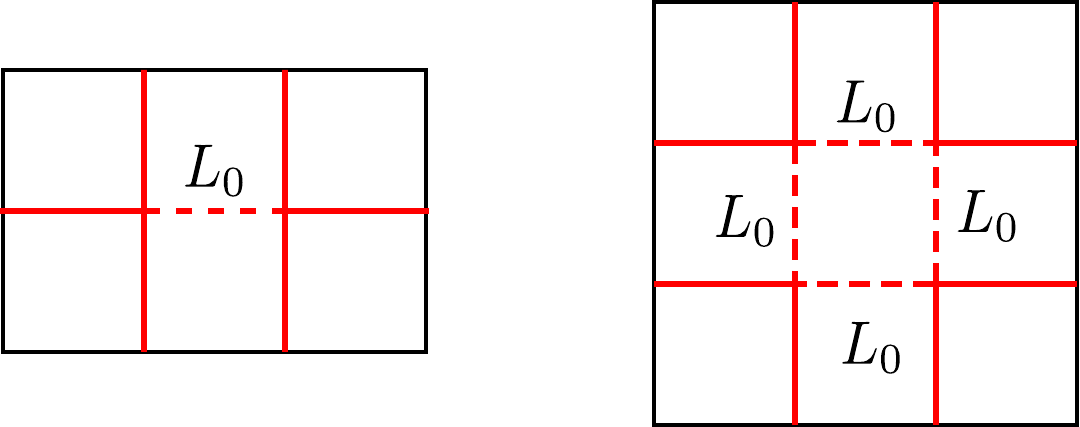}
    \caption{Left: In a 2-dimensional lattice, two vertices of degree 4 are connected by a line of length $L_0$. When exploiting the large-length limit for all the lines, but keeping $L_0$ finite, the resulting contribution from this diagram is numerically the same as the one of a 6-degree vertex. Right: 4 vertex of degree 4 are connected by lines of length $L_0$. In the large-length limit for all the lines, but keeping $L_0$ finite, the resulting contribution from this diagram is numerically the same as the one of a 8-degree vertex.}
    \label{fig:degrees}
\end{figure}

A simple question could come to mind: using the $M$-layer approach on the $ a\mathbb{Z}^D$ lattice how can one reproduce interactions with degree larger than the connectivity of the lattice, $2D$? Naively thinking, it seems impossible to map the $M$-layer construction results in every possible field theory, since the latter can always include all possible kinds of interactions that respect the symmetry of the problem. Nevertheless in the $M$-layer construction, it is possible to find vertices with interaction degrees larger than $2D$. This can be realized by connecting, for instance in $D=2$, two vertices of degree 4 by a line of length $L_0$, as shown in Fig. \ref{fig:degrees}. When exploiting the large-length limit for all the lines, but keeping $L_0$ finite, the resulting contribution from this diagram is numerically the same as the one of a 6-degree vertex. Analogously a similar procedure can be applied to realize an 8-degree vertex as depicted in Fig. \ref{fig:degrees}. A similar feature has already been found in some computations regarding the $M$-layer construction applied to the Random Field Ising model \cite{Angelini_2019}.

It is easy to generalize this argument to generate a vertex with generic degree from the $M$-layer construction.

\subsection{Soft spin variables}\label{sec:SoftSpin}
In this section we want to generalize the results, given for the Ising model with dichotomous variables, for generic soft spin variables $\sigma$, with measure $d\mu(\sigma)$. The case of dichotomous variables is recovered simply with:
\begin{equation}
    d\mu(\sigma)=\delta(1-\sigma^2)d\sigma\,.
\end{equation}
In this more general case, we should repeat the steps made in the Appendix \ref{app:transfermatrix} to compute the joint probability of two spins on an open chain, $Z_L(\sigma,\tau)$. We can introduce the cavity distribution, $Q(\sigma)$, analogously:
\begin{align}
    Q(\sigma)&=\frac{1}{Z_Q}\int d\mu(\sigma_k) e^{\beta J \sigma \sigma_k }Q^{c-1}(\sigma_k)\\
    Z_Q&=\int d\mu(\sigma_k)d\mu(\sigma)e^{\beta J \sigma \sigma_k }Q^{c-1}(\sigma_k)\,,
\end{align}
where $c$ is the connectivity of the associated RRG graph.
Here we only note that the transfer matrix
\begin{equation}
     T(\sigma,\tau)=\sqrt{Q(\sigma)^{c-2}}e^{\beta J \sigma \tau}\sqrt{Q(\tau)^{c-2}} 
\end{equation} 
is no more a $2\times2$ matrix, but an infinite dimensional operator, which has an infinite set of eigenvalues and eigenvectors:
\begin{equation}
    \frac{T(\sigma,\tau)}{\lambda_{max}}= \Vec{e}_{\lambda_{max}}(\sigma)\Vec{e}_{\lambda_{max}}(\tau)+\sum_{\substack{i=2,\dots\\|\lambda_{i+1}|<|\lambda_i|<1}}\lambda_{i}\, \Vec{e}_{\lambda_i(\sigma)}\Vec{e}_{\lambda_i}(\tau)\,,
\end{equation}
where $\lambda_2$ corresponds to $\lambda$, of the Ising variables case, and it is the one that assumes the critical value $\lambda_c=\frac{1}{2D-1}$ at the critical temperature $T_c$. The orthonormal eigenvector of $T(\sigma,\tau)$, $\Vec{e}_{\lambda_{max}}(\sigma)$, can be checked to be:
\begin{equation}\label{eq:defEigSoftSpinMAX}
    \Vec{e}_{\lambda_{max}}(\sigma)=\frac{\sqrt{Q(\sigma)^c}}{\sqrt{\int d\mu(\sigma)Q(\sigma)^c}}\,,
\end{equation}
where $\lambda_{max}=Z_Q$. On the other hand:
\begin{equation}\label{eq:normalization_softspins}
    \int d\mu(\tau)T(\sigma,\tau)\Vec{e}_{\lambda_{i}}(\tau)=\lambda_i\,\Vec{e}_{\lambda_{i}}(\sigma)\hspace{1.5cm}\text{and}\hspace{1.5cm}\int d\mu(\sigma)\Vec{e}_{\lambda_{i}}(\sigma)^2=1\,.
\end{equation}

Given this general form of $T(\sigma,\tau)$ we compute, repeating the same steps, the joint probability:
\begin{equation}\label{eq:ZL_softspins}
    Z_L(\sigma,\tau)=P_B(\sigma)P_B(\tau)+\sum_{\substack{i=2,\dots\\|\lambda_{i+1}|<|\lambda_i|<1}}\lambda_i^L\,{g_{\lambda_i}(\sigma)\,g_{\lambda_i}(\tau)}\,,
\end{equation}
where 
\begin{equation}\label{eq:P_g_soft}
    P_B(\sigma)\equiv\frac{\Vec{e}_{\lambda_{max}}(\sigma)}{\sqrt{Q(\sigma)^{c-2}}} \hspace{1.5cm}\text{and}\hspace{1.5cm} g_{\lambda_i}(\sigma)\equiv\frac{\Vec{e}_{\lambda_{i}}(\sigma)}{\sqrt{Q(\sigma)^{c-2}}}\,.
\end{equation}
Given that $|\lambda_{i+1}|<|\lambda_i|<1$ $\forall\,i\ge2$ and the fact that we want to describe the critical limit, in the regime of large lengths $L$, we neglect all the terms with $i\ge3$, obtaining, for $Z_L(\sigma,\tau)$, the same form as for the Ising variables case: $ Z_L(\sigma,\tau)\simeq P_B(\sigma)P_B(\tau)+\lambda^L\,{g_{\lambda}(\sigma)\,g_{\lambda}(\tau)}$, where we denoted $\lambda_2$ with $\lambda$. All the computations of the observables on a specific diagram go along the same way as for the Ising case. A difference with respect to the Ising variables case is that now we don't have an explicit expression for $Q(\sigma)$ and $\Vec{e}_{\lambda}(\sigma)$. This means that in computing observables we should keep track of all the constants involving the three functions $P_B(\sigma)$, $g_{\lambda}(\sigma)$ and $Q(\sigma)$. To this aim, we define a function:
\begin{equation}\label{eq:semplificazioni}
    f(x,y,z,w)\equiv \int d\mu(\sigma) \sigma^x g_{\lambda}(\sigma)^y \pb{z}{} \q{c-w}{}\,.
\end{equation}

The relevant points, useful in the following, over which $f$ is computed are:
\begin{align}
    &f(0,2,2,4) = \int d\mu(\sigma) g_{\lambda}(\sigma)^2 \pb{2}{} \q{c-4}{} = \left(\int d\mu(\sigma) Q(\sigma)^c \right)^{-1} = \int d\mu(\sigma) \pb{4}{} \q{c-4}{} = f(0,0,4,4)  \,; \label{eq:utili_1} \\
    &f(0,2,1,3) = \int d\mu(\sigma) g_{\lambda}(\sigma)^2 \pb{}{} \q{c-3}{} = \left(\int d\mu(\sigma) Q(\sigma)^c \right)^{-\frac{1}{2}} = \int d\mu(\sigma) \pb{3}{} \q{c-3}{} = f(0,0,3,3)   \,; \label{eq:utili_5} \\
    &f(0,0,1,1) = \int d\mu(\sigma) \pb{}{} \q{c-1}{} = \left(\int d\mu(\sigma) Q(\sigma)^c \right)^{\frac{1}{2}} \,; \label{eq:utili_3} \\
    &f(1,1,0,1) =  \int d\mu(\sigma)\, \sigma \,g_{\lambda}(\sigma) \,\q{c-1}{} = \int d\mu(\sigma)\, \sigma \,\Vec{e}_{\lambda}(\sigma) \,\q{\frac{c}{2}}{} \,; \label{eq:utili_4} \\
    &f(0,4,0,4) = \int d\mu(\sigma) g_{\lambda}(\sigma)^4  \q{c-4}{} = \int d\mu(\sigma)\frac{\Vec{e}_{\lambda}(\sigma)}{\q{c}{}}  \,, \label{eq:utili_2}
\end{align}
where Eqs. \eqref{eq:utili_1} and \eqref{eq:utili_5} can be verified by making use of the definitions of $P_B(\sigma)$ and $g_{\lambda}(\sigma)$ in Eqs. \eqref{eq:P_g_soft} and the normalization of $\Vec{e}_{\lambda_i}(\sigma)$ in Eq. \eqref{eq:normalization_softspins}. Now we have all the ingredients to write the results of the 2-point and 4-point connected correlation function for the Ising model with generic soft spin variables. We remark that these results are valid in the critical regime, where the lengths of the lines composing the diagrams are large. The final expressions are the same as Eqs. \eqref{eq:2point} and \eqref{eq:4point}, what change are the definitions of the parameters:
\begin{equation}\label{eq:softspincoupling}
    g\equiv -\left( \frac{f(0,4,0,4)}{f(0,0,4,4)} - 3 \right)\frac{M}{a^D\,(2D)^4}\frac{(2D)!}{(2D-4)!}\,,
\end{equation}

\begin{equation}\label{eq:softspinconstants}
    \tau\equiv \frac{M}{(2D)^2}\frac{1}{a^D} \frac{1-\lambda(2D-1) }{2D\lambda}\frac{f(0,0,1,1)^2}{f(1,1,0,1)^2}\,\qquad\text{and}\qquad \rho\equiv \frac{M}{(2D)^2}\frac{1}{a^{D-2}}\frac{1+\lambda}{4D^2\lambda(1-\lambda)}\frac{f(0,0,1,1)^2}{f(1,1,0,1)^2}\,.
\end{equation}

We note that, using the definition of the function $f$, Eq. \eqref{eq:semplificazioni}, and then the definitions of $P_B(\sigma)$, $g_\lambda(\sigma)$ and $Q(\sigma)$ for the case of Ising variables, we simply recover the results shown in the main text. On the other hand we want to stress that the definitions of the parameters, Eqs. \eqref{eq:softspincoupling} and \eqref{eq:softspinconstants}, are only valid at the critical point. This is due to the fact that we neglected all the terms with $\lambda_i$ for $i\ge3$ in Eq. \eqref{eq:ZL_softspins}, causing the definitions of the functions $f(x,y,z,w)$ to be valid only for large $L$. On the other hand these additional neglected terms don't change universal quantities, such as critical exponents.

We also notice the explicit presence of the dimensional constant $f(1,1,0,1)$ at the denominator of $\tau$ and $\rho$, as noticed in App. \ref{app:dimensionalanalysis} with the dimensional analysis.

\bibliographystyle{unsrt}
\bibliography{biblio}

\begin{thebibliography}{10}

\bibitem{Parisi1988}
Giorgio Parisi.
\newblock {\em {Statistical field theory}}.
\newblock Addison-Wesley, 1988.

\bibitem{amit2005field}
Daniel~J Amit and Victor Martin-Mayor.
\newblock {\em Field theory, the renormalization group, and critical phenomena: graphs to computers}.
\newblock World Scientific Publishing Company, 2005.

\bibitem{Zinn-Justin_2002}
Jean Zinn-Justin.
\newblock {\em {Quantum Field Theory and Critical Phenomena}}.
\newblock Oxford University Press, 06 2002.

\bibitem{FarBeyond_ch4_2023}
Tom Lubensky, Tamás Temesvári, Imre Kondor, and Maria~Chiara Angelini.
\newblock Renormalization group in spin glasses.
\newblock In {\em Spin Glass Theory and Far Beyond}, chapter~4, pages 45--67. 2023.

\bibitem{Mezard1987}
Marc M{\'e}zard, Giorgio Parisi, and Miguel~Angel Virasoro.
\newblock {\em Spin glass theory and beyond: An Introduction to the Replica Method and Its Applications}, volume~9.
\newblock World Scientific Publishing Company, 1987.

\bibitem{bethe1935statistical}
Hans~A Bethe.
\newblock Statistical theory of superlattices.
\newblock {\em Proceedings of the Royal Society of London. Series A-Mathematical and Physical Sciences}, 150(871):552--575, 1935.

\bibitem{Altieri_2017}
Ada Altieri, Maria~Chiara Angelini, Carlo Lucibello, Giorgio Parisi, Federico Ricci-Tersenghi, and Tommaso Rizzo.
\newblock Loop expansion around the bethe approximation through the m-layer construction.
\newblock {\em Journal of Statistical Mechanics: Theory and Experiment}, 2017(11):113303, nov 2017.

\bibitem{Angelini_2019}
Maria~Chiara Angelini, Carlo Lucibello, Giorgio Parisi, Federico Ricci-Tersenghi, and Tommaso Rizzo.
\newblock Loop expansion around the bethe solution for the random magnetic field ising ferromagnets at zero temperature.
\newblock {\em Proceedings of the National Academy of Sciences}, 117:2268 -- 2274, 2019.

\bibitem{angelini2018one}
Maria~Chiara Angelini, Giorgio Parisi, and Federico Ricci-Tersenghi.
\newblock One-loop topological expansion for spin glasses in the large connectivity limit.
\newblock {\em Europhysics Letters}, 121(2):27001, 2018.

\bibitem{Angelini_2021}
Maria~Chiara Angelini, Carlo Lucibello, Giorgio Parisi, Gianmarco Perrupato, Federico Ricci-Tersenghi, and Tommaso Rizzo.
\newblock Unexpected upper critical dimension for spin glass models in a field predicted by the loop expansion around the bethe solution at zero temperature.
\newblock {\em Physical review letters}, 128 7:075702, 2021.

\bibitem{baroni_2023}
Matilde Baroni, Giulia~Garcia Lorenzana, Tommaso Rizzo, and Marco Tarzia.
\newblock Corrections to the bethe lattice solution of anderson localization, 2023.

\bibitem{Rizzo_2019}
Tommaso Rizzo.
\newblock Fate of the hybrid transition of bootstrap percolation in physical dimension.
\newblock {\em Phys. Rev. Lett.}, 122:108301, Mar 2019.

\bibitem{Rizzo_2020}
Tommaso Rizzo and Thomas Voigtmann.
\newblock Solvable models of supercooled liquids in three dimensions.
\newblock {\em Phys. Rev. Lett.}, 124:195501, May 2020.

\bibitem{Fitzner_2013}
Robert Fitzner and Remco van~der Hofstad.
\newblock Non-backtracking random walk.
\newblock {\em Journal of Statistical Physics}, 150(2):264--284, jan 2013.

\bibitem{Note1}
In general, for 2-point observables that are not connected, such for example the total correlation function, additional diagrams in which the 2-points are disconnected should be considered, too. However, for connected observables such as the correlation functions the contribution of these diagrams is null.

\bibitem{Yedidia2003}
J.S. Yedidia, W.T. Freeman, and Y.~Weiss.
\newblock Understanding belief propagation and its generalizations.
\newblock In G.~Lakemeyer and B.~Nebel, editors, {\em Exploring Artificial Intelligence in the New Millennium}, chapter~8, pages 239--236. Morgan Kaufmann Publishers, January 2003.

\bibitem{bray1980renormalisation}
AJ~Bray and SA~Roberts.
\newblock Renormalisation-group approach to the spin glass transition in finite magnetic fields.
\newblock {\em J.\ Phys.\ C}, 13(29):5405, 1980.

\bibitem{moore2011disappearance}
MA~Moore and Allan~J Bray.
\newblock Disappearance of the de almeida-thouless line in six dimensions.
\newblock {\em Phys.\ Rev.\ B}, 83(22):224408, 2011.

\bibitem{angelini2015spin}
Maria~Chiara Angelini and Giulio Biroli.
\newblock Spin glass in a field: A new zero-temperature fixed point in finite dimensions.
\newblock {\em Phys.\ Rev.\ Lett.}, 114(9):095701, 2015.

\bibitem{charbonneau2019morphology}
Patrick Charbonneau, Yi~Hu, Archishman Raju, James~P Sethna, and Sho Yaida.
\newblock Morphology of renormalization-group flow for the de almeida--thouless--gardner universality class.
\newblock {\em Physical Review E}, 99(2):022132, 2019.

\end{thebibliography}

\end{document}